\documentclass[12pt,preprint]{aastex}
\usepackage{psfig}

\def\alwaysmath#1{\ifmmode{#1}\else{$#1$}\fi}

\begin{document}
\title{A 2MASS All-Sky View of the Sagittarius Dwarf Galaxy: IV. Modeling the Sagittarius Tidal Tails}

\author{
David R. Law\altaffilmark{1,2},
Kathryn V. Johnston\altaffilmark{3}, and
Steven R. Majewski\altaffilmark{2}}

\altaffiltext{1}{California Institute of Technology, Department of Astronomy, MS 105-24,
Pasadena, CA 91125 (drlaw@astro.caltech.edu)}

\altaffiltext{2}{Dept. of Astronomy, University of Virginia,
Charlottesville, VA 22903-0818 (srm4n@virginia.edu)}

\altaffiltext{3}{Wesleyan University, Department of Astronomy, Middletown, CT
(kvj@astro.wesleyan.edu)}

\begin{abstract}
M giants recovered from the 
Two Micron All-Sky Survey (2MASS) have recently been used to map
the position and velocity distributions of
tidal debris from the Sagittarius (Sgr) dwarf spheroidal galaxy entirely around
the Galaxy.
We compare this data set to both
test particle orbits and N-body simulations of satellite destruction run
within a variety of rigid Milky Way potentials
and find that the mass of the Milky Way within 50 kpc of its center should be 
$3.8-5.6\times 10^{11} M_\odot$ in order for any Sgr orbit
to simultaneously fit the velocity gradient
in the Sgr trailing debris and the apocenter of the Sgr leading debris.
Orbital pole precession of young debris and leading debris velocities
in regions corresponding to older debris
provide contradictory
evidence in favor of oblate/prolate Galactic halo potentials respectively,
leading us to conclude that the orbit of Sgr has evolved over the past few Gyr.
In light of this discrepancy, we consider constraints from the younger portions of the
debris alone within
three models of the flattening of the Galactic potential
($q =$ 0.90/1.0/1.25, i.e. oblate/spherical/prolate) in our further N-body simulations.

Based upon the velocity dispersion and width along the trailing tidal stream 
we estimate the current bound
mass of Sgr to be $M_{\rm Sgr} = 2 - 5 \times 10^8 M_{\odot}$ independant of the form of the Galactic
potential;
this corresponds to a range of mass to light ratios $(M/L)_{\rm Sgr} = 14$ - 36 $(M/L)_{\sun}$
for the Sgr core.
Models with masses in this range
best fit the apocenter of leading Sgr tidal debris
when they orbit with a radial period of roughly 0.85 Gyr and have periGalactica
and apoGalactica of about 15 kpc and 60 kpc respectively.  These distances will
scale with the assumed distance to the Sgr dwarf and the assumed depth of the Galactic
potential.
The density distribution of debris along the orbit in these models is consistent with
the M giant observations, and debris at all orbital phases where M giants are obviously present is 
younger (i.e. was lost more recently from the satellite) than the typical age of
a Sgr M giant star.
\end{abstract}


\section{INTRODUCTION}
The Sagittarius dwarf spheroidal galaxy (Sgr), discovered
only a decade ago \citep{ibata94,ibata95,ibata97}, is the most
compelling example of a satellite currently being cannibalized by the
Milky Way.  There have been numerous studies reporting the discovery
of stars and star clusters plausibly associated with debris from this
satellite, either trailing or leading it along its orbit (see Majewski
et al. 2003 --- hereafter ``Paper I'' --- for a comprehensive summary).
Using a study of faint, high-latitude carbon stars for which a significant overdensity was found to be
aligned in angular position with the projection of Sgr's orbit, Totten \& Irwin
(1998) were the first to present data that suggested that the tidal
tails of the disrupting Sgr system extend a full $360^{\circ}$ across
the sky.

The 
conclusions of the carbon star study were recently dramatically verified 
using M giants selected from the 2MASS database (Paper I).
Because Sgr is relatively metal-rich,
M giant stars are prevalent in its debris stream,
are far more common than carbon stars, and can be
easily identified to distances of more than 50 kpc within
the 2MASS database.
Moreover, the large sample of M giants in the core of Sgr itself 
permits a much more reliable distance scale 
to be derived for these stars 
than is possible 
for the carbon stars.
As a result, for the first time, primary leading and trailing tidal arms can clearly
be traced using the 2MASS M giants, with the trailing tail spanning
at least 150$^{\circ}$ across the Southern Galactic Hemisphere and the
leading tail arcing up to create a rosette orbital loop in the Northern
Galactic Hemisphere.  
Follow-up spectroscopy of Sgr-candidate stars has determined line-of-sight (i.e. ``radial'') velocities for 
Sgr M giant stars throughout the trailing tail \citep[][ hereafter ``Paper II'']{Majewski2004a} ,
and work on stars in the leading trail is in progress \citep[][ hereafter ``Paper V'']{Majewski2004b}.

 A number of groups have sought to model the Sgr --- Milky Way interaction
(e.g. Johnston, Spergel, \& Hernquist 1995, Velazquez \& White 1995,  
Ibata et al. 1997, Edelsohn \& Elmegreen 1997, Johnston et al. 1999, 
Helmi \& White 2001, G{\' o}mez-Flechoso, Fux, \& Martinet 1999, Mart{\' i}nez-Delgado et al. 2004).  
The interaction of the Sgr dwarf spheroidal with the Milky Way offers 
a sensitive probe of the shape and strength
of the Galactic potential, and also provides a nearby laboratory for
exploring the internal dynamics of satellite galaxies under the strong 
tidal influence of a parent system. 
Ibata \& Lewis (1998) made
an extensive series of simulations to match data available at the time.  Among their models,
model K6-a provides the closest match to the general morphology of the Sgr tidal tails as
mapped by M giant stars selected from 2MASS.
However, the 2MASS M giant work represents such a substantial increase in our knowledge of the
phase-space distribution of Sgr debris that a new study of the system
fully constrained by these data
is warranted.

Recently, a controversy has begun to develop over the oblate/prolate nature of the Galactic halo
as measured using Sgr tidal debris.  Helmi (2004) has presented evidence in favor of a prolate ($q =$ 1.25) halo
using Sgr leading debris velocity trends,
while in a companion paper (Johnston, Law, \& Majewski 2004, hereafter ``Paper III'') we have demonstrated that such
prolate halos fail to reproduce the observed orbital pole precession of leading vs. trailing debris, for which
oblate ($q =$ 0.90) halos best reproduce observational data.
In earlier studies, \citet{ibata01} and \citet{martinez04} determined that values
of $q \approx$ 1.0 and $q =$ 0.85 respectively best fit the available data.
In this paper, we explore whether it is possible to resolve this conflict using
a single-component (i.e. mass-follows-light) model for Sgr,
traveling  along a single orbit in a non-evolving potential.
We present the results of numerical simulations to find the best fit to the measured
positions and velocities of the M giants presented in Papers I, II, and V
(preliminary results have been presented in Law et al. 2004) while allowing orbital, potential
and Sgr internal parameters to vary.
Our aim is to constrain the current mass and orbit of Sgr as tightly as possible as 
a precursor to 
further studies in which higher order effects (such as multi-component
models for Sgr travelling along evolving orbits) are also accounted for.



In \S 2 we describe our simulation technique, and outline the properties of the observed tails that 
will be used to constrain the simulations.  
In \S 3.1 we use simple test particle simulations to
examine what range of Galactic and orbital parameters could be
consistent with Sgr debris.  
In \S 3.2 we use
the results from full N-body simulations of satellite destruction along viable orbits in
the chosen Galactic potentials to more tightly constrain the
mass and orbit of Sgr.
In \S 4 we compare our results to previous observational
and numerical work and  assess possible evolution of the Sgr orbit,
and in \S 5 we summarize our conclusions.

\section{METHOD}

\subsection{Baseline Galactic and Satellite Models}
Our simulation technique closely follows that outlined in \cite{johnston95}. The Milky Way is represented by 
a smooth, rigid potential, and Sgr by a collection of $10^5$ self-gravitating particles whose mutual interactions 
are calculated using a self-consistent field code (Hernquist \& Ostriker 1992).

A three-component model is used for the Galactic potential
and consists of a Miyamoto-Nagai (1975) disk, Hernquist spheroid, and a logarithmic halo:
\begin{equation}
        \Phi_{\rm disk}=-\alpha {GM_{\rm disk} \over
                 \sqrt{R^{2}+(a+\sqrt{z^{2}+b^{2}})^{2}}},
\label{diskeqn}
\end{equation}
\begin{equation}
        \Phi_{\rm sphere}=-{GM_{\rm sphere} \over r+c},
\label{bulgeqn}
\end{equation}
\begin{equation}
        \Phi_{\rm halo}=v_{\rm halo}^2 \ln (R^{2}+(z^2/q^2)+d^{2}).
\label{haloeqn}
\end{equation}
Following Johnston et al. (1999), we take $M_{\rm disk}=1.0 \times
10^{11}$ $M_{\odot}$, $M_{\rm sphere}=3.4 \times 10^{10}$ $M_{\odot}$,
$a=6.5$ kpc, $b=0.26$ kpc and $c=0.7$ kpc.  
In
\S 3.1 we investigate how different choices $\alpha=0.25-1.0$,
$q=0.8-1.45$, $d=1-20$ kpc and $v_{\rm circ, \odot}=180-240$ km s$^{-1}$ (the circular
speed at the Solar Circle --- $v_{\rm halo}$ in Eqn. \ref{haloeqn} was chosen to match
$v_{\rm circ, \odot}$ for a given bulge and disk contribution and adopted $d$)
affect our fit to the debris data.



Initially, the particles in our model of Sgr are distributed to generate a Plummer (1911) model
\begin{equation}
        \Phi=-{GM_{\rm Sgr,0} \over \sqrt{r^2+r_{\rm 0}^2}},
\label{PlummerEqn}
\end{equation}
where $M_{\rm Sgr,0}$ is the initial mass of Sgr and $r_{\rm 0}$ is its scale length. These particles
represent both the dark and light 
matter components of the satellite. We do not attempt to generate a 
more specific two-component model that matches Sgr's internal density and velocity distribution 
since both will evolve during the simulation. 
Rather, we explore to what extent Sgr's debris can constrain the present
global characteristics of the satellite.  These global characteristics can then be used in a more careful
consideration of the core structure in future work when better data on the core are available.

\subsection{Observational Constraints}
\begin{figure}
\plotone{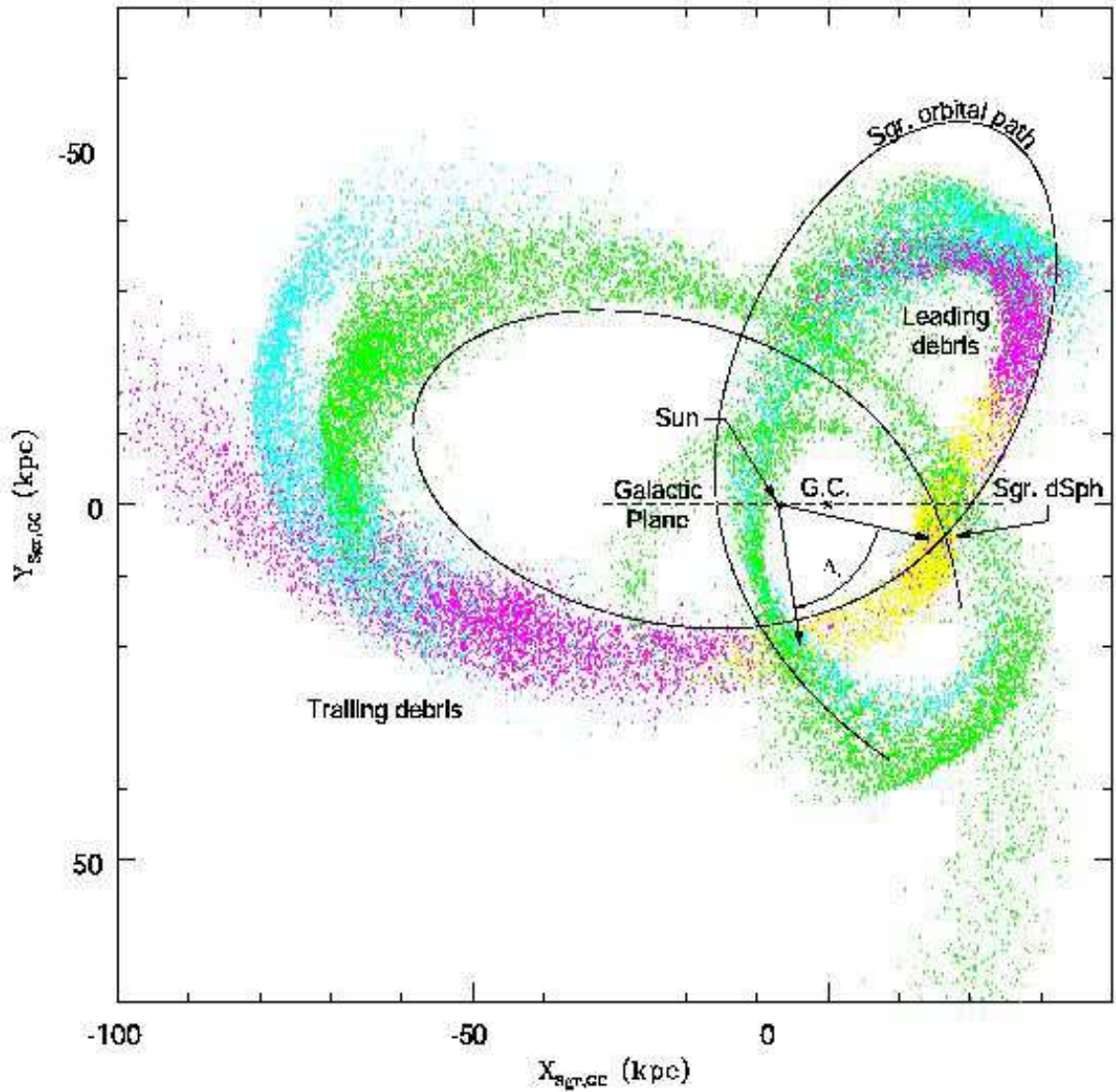}
\caption{Typical appearance of an N-body tidal debris model (colored points) in the Sgr,GC plane (this corresponding to the best 
fit $q =$ 1.0 model discussed later in \S 3.2.3).
Each color corresponds to debris lost during a single radial orbit, and 
the solid line is the projected orbit of the Sgr dwarf core.
Bold arrows define the longitudinal 
coordinate system adopted throughout this paper.
\label{xydiag}}
\end{figure}

In this paper we use the 
spherical,
Sun-centered, Sgr-coordinate system\footnote{C++ code to convert from standard
Galactic coordinate systems to the Sgr longitudinal coordinate system can be obtained from the World Wide Web at
http://www.astro.virginia.edu/{$\sim$}srm4n/Sgr/} defined in Paper I, 
since this is the coordinate system in which
satellite debris is 
observed and therefore can be compared to simulations most clearly.
The zero-plane of the latitude coordinate 
$B_{\odot}$
coincides with the 
best-fit great circle defined by Sgr debris, as seen from the Sun.
The longitudinal coordinate
$\Lambda_{\odot}$ is zero in the direction of
the Sgr core and increases along
the Sgr trailing debris stream, i.e. away from the Galactic plane.
Figure  \ref{xydiag} shows a representative N-body simulation of the Sgr dwarf in Cartesian
$X_{\rm Sgr,GC}$, $Y_{\rm Sgr,GC}$  coordinates 
(see Paper I for the definition of the Sgr,GC and Sgr,$\odot$ coordinate systems), and illustrates
the orientation of the spherical coordinate system with respect to the Galactic Plane.
The colors of the simulated data used in this and other figures in this paper represent different debris ``eras'', 
i.e. orbits (denoted as one 
apoGalacticon to the next apoGalacticon)
on which the debris was stripped 
from the satellite.
Yellow points represent debris stripped from the satellite since apoGalacticon about 0.5 Gyr ago, while magenta, 
cyan, and green points represent 
debris stripped from the dwarf 2, 3, and 4 orbits ago respectively.  
Note that while each color represents debris unbound from the satellite between two successive apoGalactic passages,
the majority of debris of each color is released during the corresponding periGalactic passage.
This color scheme allows us to discriminate readily between different wraps of tidal debris, and is also useful for
determining the expected age of debris at any given point along the tidal stream.

The observed position and radial velocity data for Sgr M giants 
(Papers I, II, and V) provide strong constraints on the 
orbit of the Sgr dwarf.
Most other Sgr detections around the sky fall within
the M giant-traced tails (see Fig. 17 of Paper I); therefore 
we compare our models to the M giants alone because they offer the most consistent,
wide-ranging map of Sgr debris, and at the same time encompass the previous detections.
We compare our results to the recently announced SDSS detections \citep{newberg03}
in \S 4.1.

We define eleven observed properties that we adopt as constraints on our simulated Milky Way --- Sgr system:
\begin{enumerate}
	\item The model Sgr dwarf should be located at $(l,b) = (5.6^{\circ},-14.2^{\circ})$ (Paper I).
	\item The line-of-sight velocity\footnote{All velocities are given in the Galactic Standard
of Rest (GSR) frame.} of the model dwarf should be $v_{\rm los,Sgr} =$ 171 km s$^{-1}$ (Ibata et al. 1997).
	\item Sgr debris should be aligned with the plane passing through the Sun having a
pole  $(l,b)=(273.8^\circ,-13.5^\circ)$  (Paper I).
	\item The average heliocentric distance for Sgr leading debris at apoGalacticon ($d_{\rm avg}$) in the Northern Galactic 
Hemisphere should be 42 $\times (D_{\rm Sgr}$/24 kpc) kpc, where $D_{\rm Sgr}$ is the assumed distance to 
Sgr (which sets the distance scale of the M giants and is taken to be 24 kpc in Paper I). 
	\item Radial velocities along the trailing stream from $\Lambda_{\odot} = 25^{\circ}$ - $140^{\circ}$  
should match data presented in Paper II.
	\item Radial velocities along the leading stream from $\Lambda_{\odot} = 230^{\circ}$ - $330^{\circ}$  
should match data presented in Paper V.
	\item The leading and trailing debris tails should define two distinct planes with poles offset
from each other by $\sim$ 10 degrees (Paper III).
	\item The physical width of the trailing debris stream perpendicular to the orbital plane 
should be consistent with M giant observations (i.e. have a projected spatial dispersion 
$\sigma_{Z_{\rm Sgr,\odot}} \approx 2.0$ kpc).
	\item The average radial velocity dispersion along the trailing stream
	        from $\Lambda_{\odot} = 25^{\circ}$ - $90^{\circ}$
		should match the dispersion found for M giants in Paper II ($\sigma_v$ = 10.0 km s$^{-1}$).
	\item The model debris to which the M giant data are matched should be younger (i.e. have left Sgr more recently) 
than a typical Sgr M giant age (2-3 Gyr; see Paper I).
	\item There should be a break in the surface density of trailing debris at $\Lambda_{\odot} \sim 20^\circ$, 
which has previously been interpreted (Mateo et al. 1998, Paper I) to correspond to 
the transition between debris lost on the current pericentric passage and that lost on the previous passage.
\end{enumerate}

\section{RESULTS}

Table \ref{param.table} outlines the Galactic and satellite parameters varied to
produce a model that fits the constraints detailed above.  Rather than
run lengthy N-body simulations to randomly search for a global minimum
in this degenerate, multi-dimensional parameter space (8 of which are allowed
to vary), a more
efficient, multi-step approach was taken to converge to the best fit
to the observational data, relying on physical insight gained both
from analytical descriptions of debris dispersal (Tremaine 1993,
Johnston 1998, Helmi \& White 1999, Johnston, Sackett \& Bullock 2001)
and from previous modeling of Sgr by the authors (Johnston, Spergel,
\& Hernquist 1995, Johnston et al. 1999) and other groups (Velazquez
\& White 1995, Ibata et al. 1997, Edelsohn \& Elmegreen 1997, G{\'
o}mez-Flechoso, Fux, \& Martinet 1999, Mart{\' i}nez-Delgado et
al. 2004).  These studies have found that while there is a systematic
distance offset for leading/trailing debris inside/outside the orbit
of the Sgr dwarf (a reflection of the debris moving to more/less
tightly bound orbits --- see Fig. \ref{xydiag}) the line-of-sight
velocity remains approximately aligned with that of the satellite's
orbit at all orbital phases.  Hence in \S3.1 below we are able to
eliminate a wide range of orbits in a variety of Galactic potentials
through test-particle integrations alone: We use constraint 4
as an upper limit on a possible orbit's apocentric distance and
examine how well the line-of-sight velocities along the orbit match
the data in constraints 5 and 6.  This technique allows us to find
reasonable values for all of the free parameters listed in Table
\ref{param.table} except for Sgr's current mass.  In \S 3.2 we describe 
full-scale simulations of the destruction run for satellites of various
masses along the orbits and in the potentials selected in \S3.1.

\subsection{Galactic Parameters}

 
\subsubsection{Varying initial conditions for test particle orbits}

We assume Sgr's current angular position, line-of-sight velocity and
direction of proper motion to be fixed by constraints 1, 2 and 3
respectively, and adopt an amplitude for the motion of Sgr
perpendicular to our line-of-sight ($v_{\rm tan}$) somewhere within
$\pm 3$ times the error bars on the \citet{ibata01} measurement of
$280\pm20$ km s$^{-1}$.  The Sgr velocity and position relative to the Sun are then tranformed to
Galactocentric coordinates to provide initial conditions for the test
particle orbits, assuming some values for the Solar distance from the
Galactic center ($R_\odot$) and from Sgr ($D_{\rm Sgr}$).  (Note that
changing $D_{\rm Sgr}$ from the assumed value of 24 kpc scales the
distances to all of the Sgr M giants by the same fractional amount,
since these distances are estimated from a color-apparent magnitude
relation derived from M giants in Sgr's core --- Paper I).  These
orbits are then integrated backwards and forwards in time in the
chosen Galactic potential (see \S3.1.2) and the quality of fit of the
orbital path to the M giant position and velocity data quantified (as
described in \S3.1.3).

\subsubsection{Varying the Galactic potential --- parameter and model choices}

 We anticipate that Sgr's debris will tell us something about the
contours of the gravitational potential in the region that its orbit
explores ($\sim$ 10 - 50 kpc). Hence we do not vary all parameters in
equations (1) - (\ref{haloeqn}), but instead hold the bulge component
fixed and explore the effect of changing the contribution of the disk to
the rotation curve through the parameter $\alpha$, as well as the
radial length scale, flattening, and overall depth of the halo potential
respectively through the parameters $d$, $q$ and $v_{\rm circ, \odot}$.

As a final check on the generality of our results, we repeat our
experiments with the halo component replaced by models of the form
proposed by \citet{NFW96} --- hereafter referred to as NFW models.  In
this case, the flattening is introduced in the density $q_{\rho}$,
rather than potential contours, and the approximate form of the
potential is adopted from \citet{jing02}.  To explore a similar effective range
in $q$ and radial gradient as the logarithmic models, $q_{\rho}$
and $d$ (the length scale of the NFW potential) are chosen from a
wider range than for their logarithmic counterparts --- in particular,
the range $5<d<100$ kpc was explored because this encompasses the range
of scale lengths (of order tens of kpc) found for dark matter halos of
similar mass-scale to the Milky Way in cosmological models of
structure formation at the current epoch \citep[e.g.][]{eke01}.  The
mass scale of the NFW potential is then constrained to match the
adopted $v_{\rm circ, \odot}$.


\subsubsection{Quantifying the fit of an orbit to the data.}

A guideline for assessing the goodness of fit of an orbit to the
positional data is that the maximum heliocentric distance observed for
the leading debris ($D_{\rm debris}$, constraint 4) must be
systematically less than that of the orbit of the Sgr core, $D_{\rm
max}$ --- i.e. $D_{\rm max}/D_{\rm debris}>1$.  We can also find an
upper limit to this ratio since we expect the size of this offset to
scale as $\Delta R \propto R(M_{\rm Sgr}/M_{\rm Gal})^{1/3}$, where
$M_{\rm Gal}$ is the mass of the Milky Way enclosed within the
pericenter of the orbit \citep{johnston01}.  For example, if we take
this limit as $D_{\rm max}/D_{\rm debris}<1.5$ then we might expect to
cover all models with $M_{\rm Sgr}/M_{\rm Gal}<0.125$ --- i.e. Sgr masses up
to 10\% of the mass of the Milky Way.  Since the
internal dispersion measured for Sgr \citep[11 km
s$^{-1}$,][]{ibata95} suggests a mass far less than this we take
$1<D_{\rm max}/D_{\rm debris}<1.5$ as a generous range for considering
an orbit apogalacticon distance acceptable. Orbits with apogalactica outside
this range are immediately rejected.

\begin{figure}
\plotone{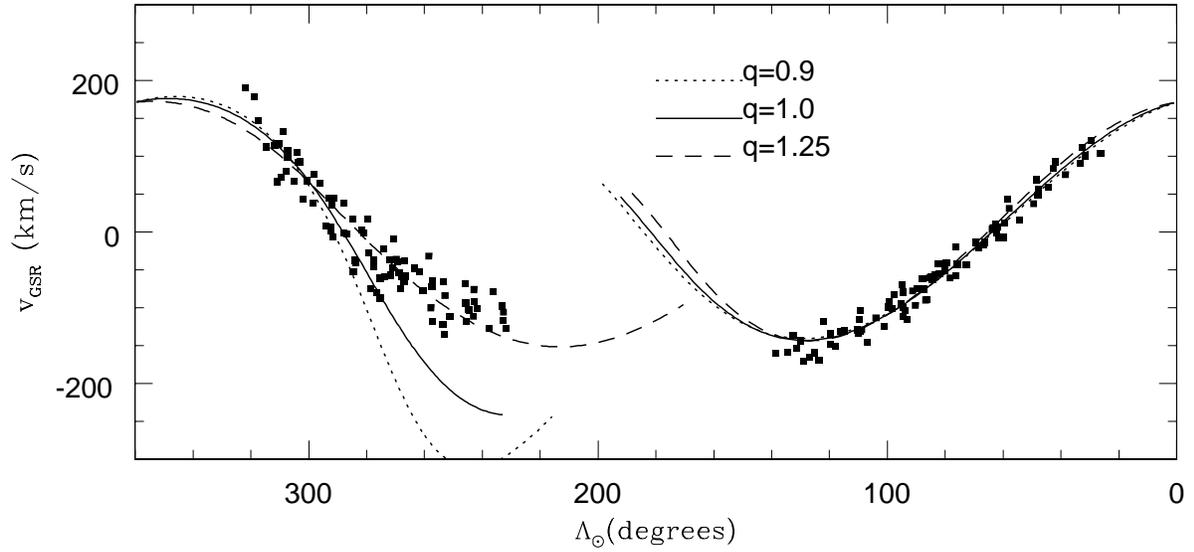}
\caption{Square points show the selected velocity data in leading and trailing arms that
represent the general trend and dispersion of Sgr debris in these regions. 
Solid/dashed/dotted curves show ``best'' (as defined in \S3.1.3) orbits selected to fit 
the trailing data alone in the final potentials adopted with the specified $q$ in \S3.1.5C }
\label{orb.fig}
\end{figure}

We next quantify the fit of orbits that are not already rejected 
 to the trailing and leading velocity data (constraints 5 and 6)
through the parameters $\chi_{\rm trail}$ and  $\chi_{\rm lead}$ :
\begin{equation}
\chi_{\rm A}^2={1 \over N_{\rm A}}\sum_{i=1}^{N_{\rm A}}{[v_{\rm M giant, A}(\Lambda_{\sun})-
v_{\rm orb}(\Lambda_{\sun})]_i^2\over{\sigma^2_{\rm A}}}
\label{khia}
\end{equation}
where ``A''  represents the observed data set being considered (i.e. ``lead'' or ``trail''), $N_{\rm A}$ is
the number of M giants in the data set,
$v_{\rm M giant}(\Lambda_{\sun})$ is the velocity of an M
giant at $\Lambda_{\sun}$ and $v_{\rm orb}(\Lambda_{\sun})$ is the
velocity of the orbit at this $\Lambda_{\sun}$.  The data compared to
in the leading portion of the debris 
are selected by fitting a
3rd order polynomial to the full data set of velocities as a
function of $\Lambda_{\odot}$ 
in the range $230^\circ<\Lambda_{\odot}<330^\circ$. Outliers from the main trend are thrown out using
a 2.5-$\sigma$ iterative rejection technique until convergence is
reached and the weight $\sigma^2_{\rm lead}$
calculated as the dispersion of the velocities of this 
final set of $N_{\rm lead}$ stars about the best-fit polynomials.
The process is then repeated for stars in the region
$25^\circ<\Lambda_{\odot}<140^\circ$ most sensitive to the
trailing debris. The selected stars in both regions are plotted as black squares in Figure \ref{orb.fig}.
Clearly, these data sets are not intended to represent a complete sample of Sgr stars, but
rather as a guide to the general trends of velocities and dispersion in these regions.

We also express these quantities as a single parameter to measure the combined 
goodness-of-fit:
\begin{equation}
\chi=\sqrt{\left(\chi^2_{\rm trail}+\chi^2_{\rm lead}\right)/2}.
\label{khi}
\end{equation}


Note that since test particle orbits only serve as an indication of where the
debris should lie, we do not simply search for the parameters corresponding to the minima of these
quantities: for example, we do not consider 
a difference of order $\Delta \chi < 0.1$  (corresponding 
to average systematic offsets $\sim 1-2$ km s$^{-1}$  --- very much less than the dispersion in the data) 
between the fit to two different orbits to be very significant. Rather, we use more
extreme differences to rule out or favor broad regions of parameter space.

\subsubsection{Combined constraints from leading and trailing velocity data}

Although the velocity trends in the leading 
debris (constraint 6) appear to strongly favor Galactic models with prolate ($q>1$) halo components 
\citep{helmi04}, we have shown in Paper III that 
the direction of the precession of debris orbits (as measured by the 
offset in the poles of best-fit planes to leading and trailing debris --- constraint 7) strongly
favors models with oblate halos
since prolate models induce precession in the {\it opposite} sense to that observed.
Because no other adjustment to the potential can change the
fundamental sense of precession in prolate {\it vs} oblate potentials, 
we restrict ourselves to asking whether we can resolve this contradiction between
the implications of constraint 6 and constraint 7 by revisiting the fit to the velocity 
and distance data alone over a much wider range of parameter
space than has been considered previously. 
The aim is to examine whether there are any circumstances in which 
an orbit in an oblate potential can  be found
that can fit all the constraints at once.

\begin{figure}
\plotone{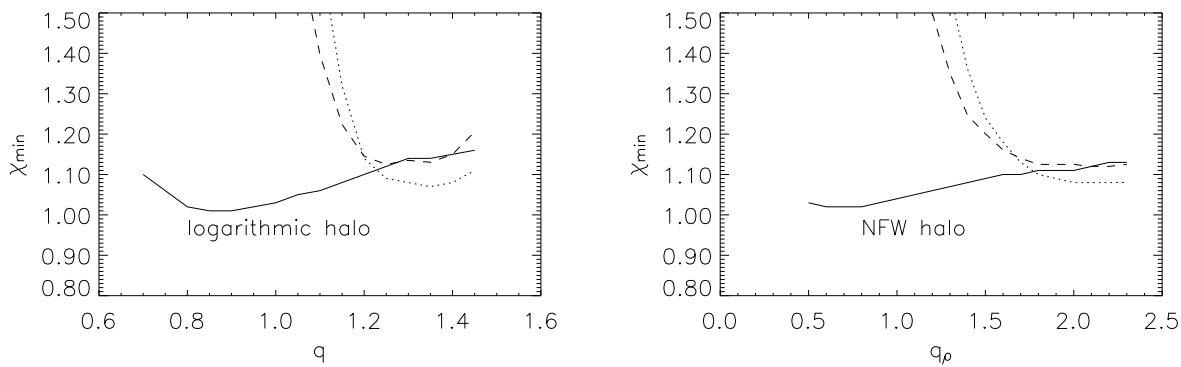}
\caption{Minimum values of $\chi_{\rm trail}$ (solid lines, equation
\ref{khia}), $\chi_{\rm lead}$ (dotted lines, equation \ref{khia})
and $\chi$ (dashed lines, equation \ref{khi}) as a function of $q$ (in
potentials with logarithimic halo components --- left hand panel) or
$q_\rho$ (in potentials with NFW halo component --- right hand panel) when
all other parameters are varied freely.}
\label{khi.fig}
\end{figure}

Figure \ref{khi.fig} plots the minimum values of $\chi_{\rm trail}$ (solid lines), 
$\chi_{\rm lead}$ (dotted lines) and $\chi$ (dashed lines)
obtained as a function of $q$ (left hand panel, logarithmic halo model)
or $q_\rho$ (right hand panel, NFW halo model)
when all other parameters are allowed to vary freely within
the ranges outlined in Table \ref{param.table}.
The solid lines show that the trailing velocity data have
a slight preference for models with oblate halos, although the difference $\Delta \chi_{\rm trail}\sim 0.1$
between the minima for models with $q<1$ and $q>1$ is not sufficiently large that
we can confidently rule out prolate models with test particle orbits alone, since it corresponds to
a velocity offset much less than the dispersion in the data.
In contrast, the dotted lines show that leading velocity data strongly prefer
prolate halo models, to such an extent that this preference dominates the combined
$\chi$ (dashed lines). These results are the same for the logarithmic and NFW models.

Overall, we conclude that we cannot find a single orbit in a static
potential model that simultaneously fits the velocity data in the
trailing data together with the sense of precession suggested by the
offset of the planes of the leading {\it vs} trailing data.

\subsubsection{Constraints from trailing velocity data alone}

The exciting implication of the conclusion of the previous section 
--- that no {\it single} orbit and/or potential can fit all the data --- is
that some evolution of Sgr's orbit has occurred over the time since
debris in the leading portion of the streamer, furthest in $\Lambda_\odot$ from Sgr,
was released. We discuss
some possible culprits for this orbital evolution in \S4.3, but defer
a detailed investigation of these effects for future work. For the
remainder of this study, we narrow our present analysis to concentrate 
on the younger
portions of the debris, lost within the last 1-2 orbits, where (1) the
effect of orbit evolution is negligible, (2) the modelling can 
be acheived with the fewest free parameters,
and (3) the interpretation of the data is less ambiguous.
The goal is to ask what the younger debris alone
can tell us about the Galactic potential and Sgr's current mass and
orbit.
These results can subsequently be used 
as starting points for studies that use the older debris to examine 
higher order effects such as
orbital evolution, evolution of the potential, and/or multi-component models for Sgr.

We expect debris in the trailing streamer in the range explored by
the velocity data to be roughly the same age as that in the early
parts of the leading streamer to about the first apocenter \citep[as
demonstrated by][and see also \S3.2 below]{helmi04}.  In these
regions, the velocity data can be similarly fit by both oblate and
prolate potentials (as demonstrated by the solid lines in Figure \ref{khi.fig}), 
and there is no significant offset in the orbital
poles between the leading and trailing components.  
Hence, we now drop
constraints 6 and 7 on our models since these were derived from regions where orbit evolution
could be significant.
We continue our discussion of
test-particle constraints on the Galactic potential and our position
relative to the Galactic center and Sgr using the condition $1<D_{\rm
max}/D_{\rm debris}<1.5$ and examining $\chi_{\rm trail}$ alone.
 
In order to sort through our large parameter space, we first look at parameters
that do not appear to be strongly constrained by the data and fix
reasonable values for those (see discussion in A. and B. below) before
going on to look at preferred ranges for the remaining parameters (in C.).
 
\subsubsubsection{A. Implications for distance scales}

\begin{figure}
\plotone{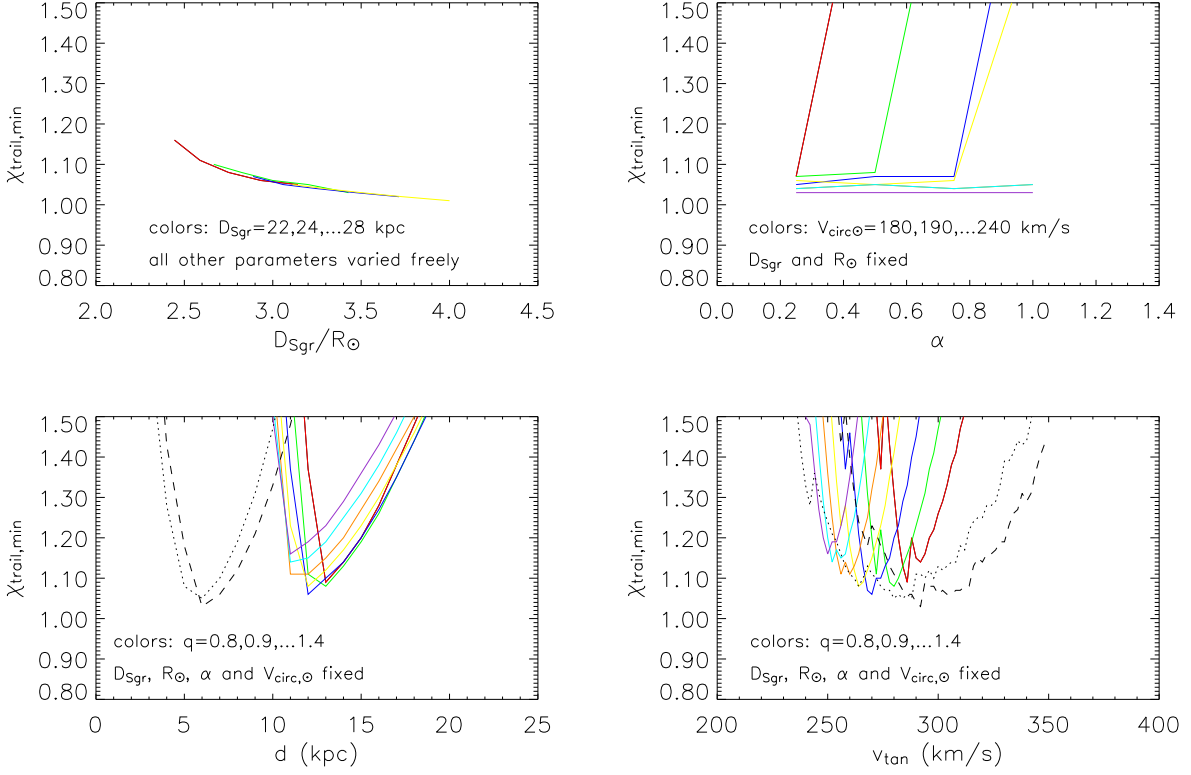}
\caption{Projections of results in our 7-dimensional parameter space 
onto two-dimensions for experiments with logarithmic halo components.
One axis of the plane is plotted along the $x$-axis and the other
represented by the different colored lines --- 
the upper
label in
each panel gives the second dimension explored with the numbers
corresponding in sequence to red/green/blue/yellow/orange/light
blue/violet lines. 
(Note: in some panels certain colors appear to be missing in the sequence because the
lines are overplotted on top of one another.)
The $y$-axis shows the minimum $\chi_{\rm trail}$ in the
illustrated plane when:
 (i) {\it upper left hand panel ---}  all other parameters vary freely;  (ii) {\it upper right hand panel ---}  
$D_{\rm Sgr}=24$ kpc and $R_\odot$ = 7 kpc
and (iii) {\it lower panels ---}
$D_{\rm Sgr}=24$ kpc, $R_\odot$ = 7 kpc, $v_{\rm circ,\odot}=$ 220 km s$^{-1}$ and $\alpha=1$ .
Dashed black curves in the lower panels outline where colored curves would fall
with same fixed parameters but $v_{\rm circ,\odot}=$ 240 km s$^{-1}$.
Dotted black curves outline the location for $v_{\rm circ,\odot}=$ 220 km s$^{-1}$ and $\alpha=0.5$. }
\label{ktrail_log.fig}
\end{figure}

\begin{figure}
\plotone{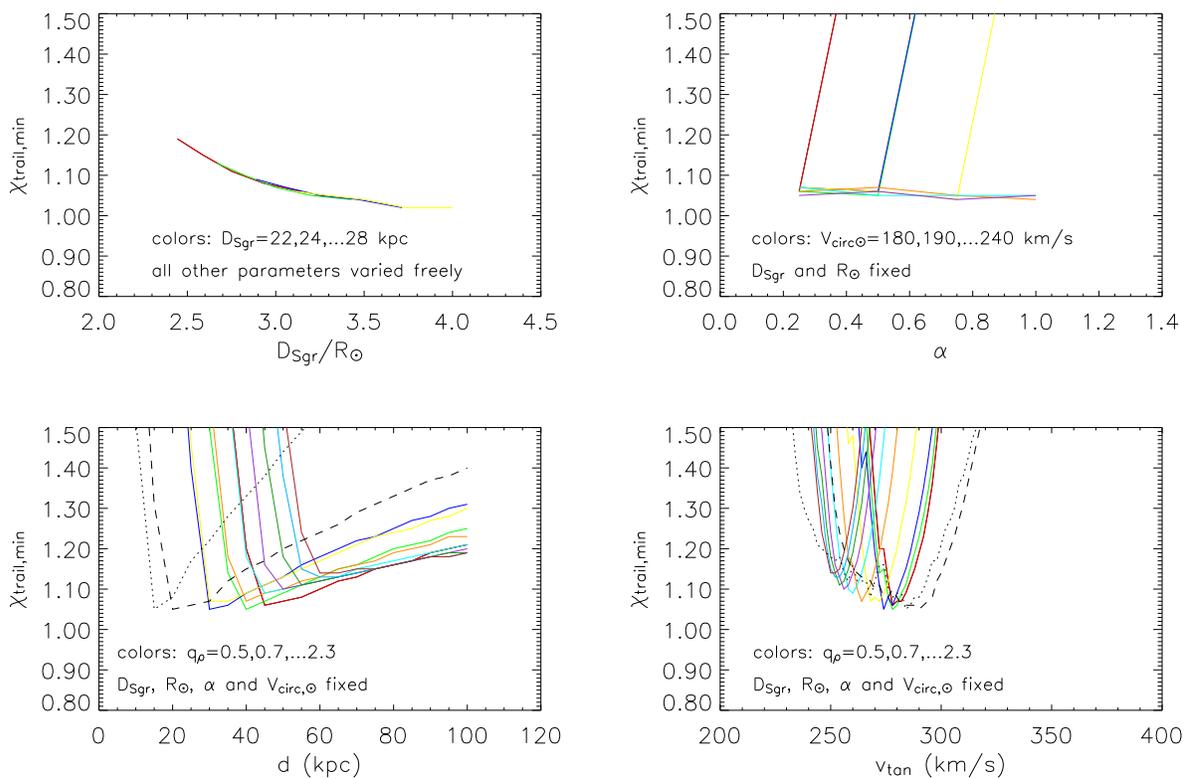}
\caption{As Figure \ref{ktrail_log.fig} but for models with NFW halo components. In this case
$v_{\rm circ,\odot}$ is held fixed at 230 km s$^{-1}$ in the colored and dotted black lines in
the lower panels. All other fixed quantities in the lower panels are the same.}
\label{ktrail_nfw.fig}
\end{figure}

In the upper left hand panels of  Figures
\ref{ktrail_log.fig} (for logarithmic halo experiments) and \ref{ktrail_nfw.fig} (for NFW halo
experiments) we project results in our 7-dimensional parameter space
onto the 2-dimensions of $D_{\rm Sgr}$ and $D_{\rm Sgr}/R_{\odot}$  
by plotting the minimum value of $\chi_{\rm trail}$ at each point in this
plane when all other parameters are allowed to vary freely.
The plots reveals a preference for larger values of the ratio $D_{\rm Sgr}/R_\odot$, with
the absolute scale (as set by $D_{\rm Sgr}$) being arbitrary.
For consistency with the distance scales adopted earlier in Paper I we choose to take $D_{\rm Sgr}=24$ kpc
\citep[which also lies within the 2-$\sigma$ error bars of the recent measurement by][]{monaco04} and
set $R_\odot=7$ kpc.
So long as $D_{\rm Sgr}/R_{\odot}\sim 3.4$ we expect all subsequent 
results involving distances (e.g. scale-length of the halo $d$, or predicted distances to debris)
can be scaled by whatever value $D_{\rm Sgr}$ is assumed in a given study.

\subsubsubsection{B. Implications for the Galactic rotation curve}

With $D_{\rm Sgr}=24$ kpc and $R_\odot=7$ kpc fixed, 
the upper right hand panels of  Figures \ref{ktrail_log.fig} and \ref{ktrail_nfw.fig} project
the remaining five-dimensions of parameter space onto the $v_{\rm circ, \odot}$ - $\alpha$
plane. For high enough $v_{\rm circ, \odot}$ there is no preference for a particular $\alpha$,
but models with lower $v_{\rm circ, \odot}$ are inconsistent with heavier Galactic disks
(i.e. higher $\alpha$). 

\begin{figure}
\plotone{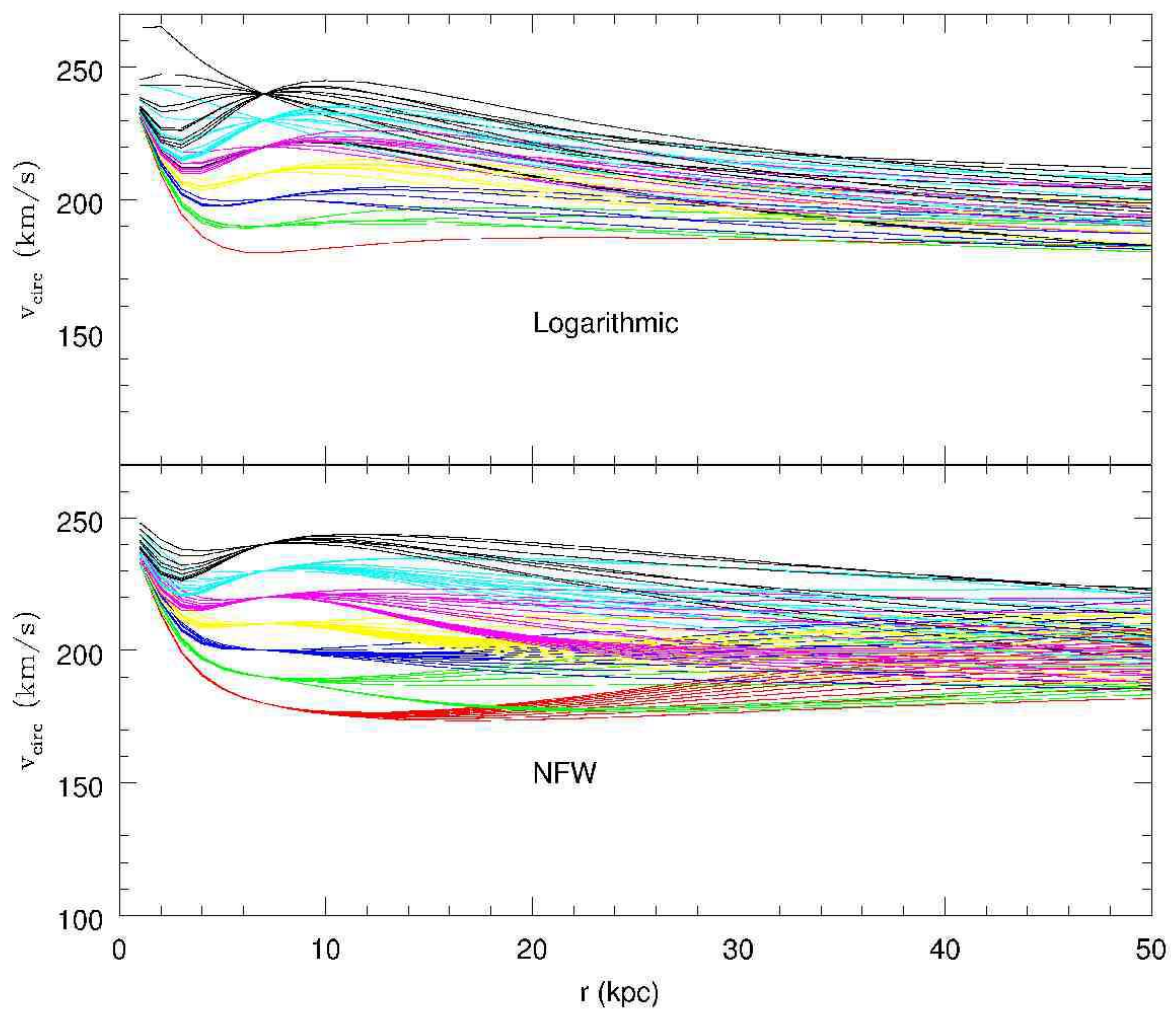}
\caption{Rotation curves for all models with logarithmic (upper panel)
or NFW (lower panel) halo components in which orbits can be found that
satisfy both $1<D_{\rm max}/D_{\rm debris}<1.5$ and $\chi_{\rm
trail}<1.1$. Colors black/cyan/magenta/yellow/blue/green/red correspond to potentials with
$v_{\rm circ, \odot}=240/230/220/210/200/190/180$ km s$^{-1}$ at $R_\odot=7$ kpc.}
\label{rot.fig}
\end{figure}

Figure \ref{rot.fig} offers some clue as to why this is the case by
plotting rotation curves for only those potentials in which orbits
with $\chi_{\rm trail}<1.1$ could be found.  These are very flat out
to large radii for all models, with circular velocities at 50 kpc in
the range 180-220 km s$^{-1}$ (which corresponds to enclosed masses for the
Milky Way at these radii of $3.8-5.6 \times 10^{11}M_\odot$ --- in
effect, Sgr debris velocities are now providing additional evidence
for the existence of a dark matter halo to the Milky Way).  If enough
of the contribution $v_{\rm circ,\odot}$ is provided by the disk, then
the remaining halo component is simply not massive enough to support such an
extended flat rotation curve. (Larger mass halos could be built by allowing $d$
an even wider range, but these models would [i] have rising rotation curves at the
Solar Circle; and [ii] be inconsistent with scale lengths measured for
Milky Way-sized dark matter halos formed in cosmological models of structure formation
--- see Eke, Navarro \& Steinmetz, 2001).

Since no values of $\alpha$ and $v_{\rm circ, \odot}$ are at this point clearly preferred,
we adopt $\alpha=1.0$ and $v_{\rm circ, \odot}=220$ km s$^{-1}$.

\subsubsubsection{C. Summary of parameter choices and conclusions}

The colored lines in the lower panels of Figures \ref{ktrail_log.fig}
and \ref{ktrail_nfw.fig} demonstrate that, with $R_\odot=7$ kpc,
$D_{\rm Sgr}=24$ kpc, $\alpha=1$ and $v_{\rm circ, \odot}=220$ km s$^{-1}$
fixed, particular values for $d$ (which determines the radial gradient
of the potential and hence the shape of the rotation curve) and
$v_{\rm tan}$ (which determines the scale of the orbit within this
potential) are quite strongly preferred, with only a mild dependence
on $q$.  Hence we perform full N-body simulations in potentials with
logarithmic halos in which
$q=0.9/1.0/1.25$, $d=13/12/11$ kpc (from the minima in the lower left
hand panels) and $v_{\rm tan}$ in the range $\pm 20$ km s$^{-1}$ around $280/270/254$ km s$^{-1}$.
All three values, $q=0.9/1.0/1.25$, are considered since all represent 
equally viable fits to the younger debris.

Clearly, our choices are not unique. 
The black curves in the lower panels of Figures \ref{ktrail_log.fig}
and \ref{ktrail_nfw.fig} outline where the colored lines would fall if
all other parameter choices were the same but
$v_{\rm circ, \odot}=240$ km s$^{-1}$ (dashed lines) or $\alpha=0.5$ (dotted lines).
In both cases, the scale-length changes significantly in order to maintain
the necessary flatness of the rotation curve, and $v_{\rm tan}$ is similarly affected.

In addition, our decision to use logarithmic halos rather than NFW halos is arbitrary, since 
Figures \ref{ktrail_log.fig} and \ref{ktrail_nfw.fig} reveal no preference for either form of the 
potential, but rather more generally indicate that any model that generates a flat rotation
curve out to 50 kpc will suffice.
We anticipate that data exploring even larger distances from the Galactic center will be
able to address whether an NFW (with a falling rotation curve in this region)
or logarithmic potential is more appropriate.

Despite these multiple minima in parameter space, we are able to reach
some general conclusions at this point: (i)
Sgr debris data prefers models with large values of $D_{\rm Sgr}/R_\odot$
and flat rotation curves out to 50 kpc, and (ii)
with all other parameters fixed,
Sgr orbits in prolate halos will have systematically lower $v_{\rm tan}$ than in spherical 
or oblate halos.
These conclusions offer a tantalizing glimpse of how Sgr debris might be used to map
out the Galactic potential on large scales once parameters such as $R_\odot$
and $v_{\rm tan}$
are known with more certainty.

\subsection{Sagittarius'  Properties}

Using the Galactic parameters determined in \S 3.1 above, 
we now perform fully self-consistent N-body 
simulations to refine the estimates obtained in \S 3.1 of Sgr's orbital velocity and to determine the mass of the dwarf.
These simulations
 follow the evolution of satellites with a range of initial masses and physical scales (varied through the 
parameters $M_{\rm Sgr, 0}$ and $r_0$ in Equation [\ref{PlummerEqn}]) along a small range of 
plausible orbits within the three models of the Galactic potential ($q =$ 0.9/1.0/1.25)\footnote{We adopt the convention
of stating values derived in each of these potentials for oblate/spherical/prolate cases, respectively.} discussed in \S 3.1.5C.

In \S 3.2.1 we find the mass of Sgr (independent of $r_0$) that best fits constraints 8 and 9 in each of these three models of the Galactic potential,
and demonstrate that this best-fit mass is common to all three cases.
Fixing the satellite mass to this best-fit value, we refine our estimate for Sgr's tangential velocity using constraints 4 and 5
in \S 3.2.2 and summarize the properties of our best-fit models in \S 3.2.3.

\subsubsection{Constraining the Mass of the Sgr Dwarf}

While we do not attempt to model the Sgr core in detail, we are nonetheless able to constrain its current total mass under the 
assumptions that the dwarf is roughly spherical and non-rotating.
Motivated by previous work (e.g., Johnston, Hernquist, \& Bolte 1996, Johnston 1998) 
we expect that debris width (constraint 8) and velocity dispersion (constraint 9) at a given orbital phase 
primarily reflect the mass within the tidal radius of the satellite on the orbit immediately prior to that debris 
becoming unbound, and that they do not depend strongly on  
the internal structure of the satellite (in our case parameterized by the scale length of the 
initial Plummer model).
For the same reasons, we do not expect that our results are strongly sensitive 
to the particle distribution we have adopted. 
We do expect the internal orbital distribution will independently affect debris morphology, but do not address that issue in this paper.

To compare the simulations to the data constraints, we calculate the average radial velocity dispersion $\sigma_{\rm v}$ and the average
dispersion of distances perpendicular to the Sgr plane $\sigma_{\rm Z_{\rm Sgr,\odot}}$ in the trailing tail for M giant data and our numerical
simulations.  We do not consider leading debris in obtaining our mass estimates since only our prolate halo model successfully matches
the {\it bulk} trend of leading debris, while all three halo models reproduce the trailing debris trend.
$\sigma_{\rm v}$ is calculated in the range $\Lambda_{\odot} = 25^{\circ}$ - $90^{\circ}$ for consistency with the velocity dispersion
analysis presented in Paper II, while $\sigma_{\rm Z_{\rm Sgr,\odot}}$ is calculated in the range
$\Lambda_{\odot} = 60^{\circ}$-$120^{\circ}$ since this range of debris longitudes
is one for which
all Sgr stars in the sample\footnote{This sample is drawn directly from the 2MASS database with the selection criteria 
$E(B-V) < 0.555$, $1.0 < J-K < 1.1$, $|Z_{\rm Sgr,\odot}| < 5$, $Z_{\rm GC} < 0$, and 13 kpc $< d_{\ast} <$ 40 kpc.}
are at a similar distance $d_{\ast}$ from the Sun (this minimizes artificial width inflation on the sky
due to differential
distance errors) and is also in a region of the Galaxy where sample contamination 
by Milky Way disk stars is negligible.

\begin{figure}
\plotone{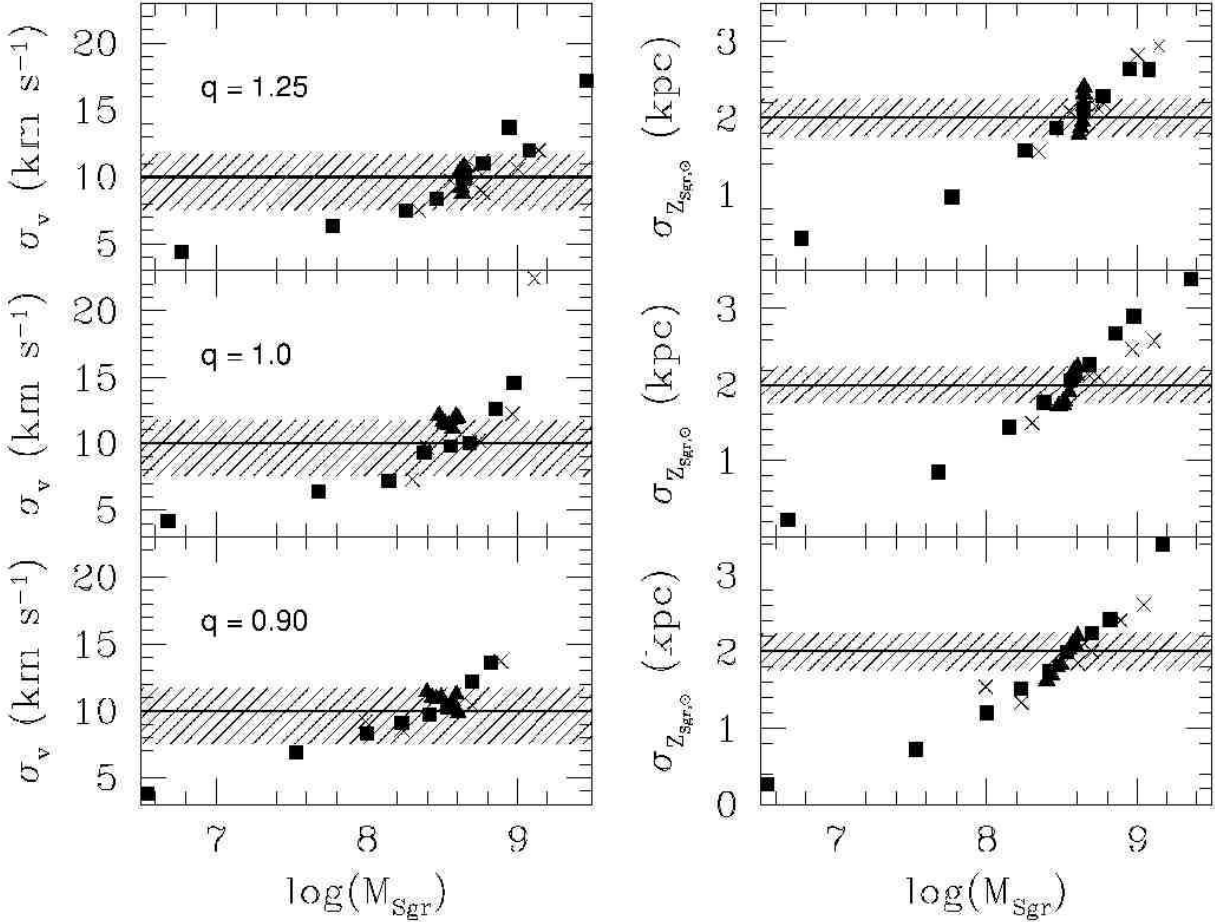}
\caption{The 
velocity dispersion of trailing satellite debris ($\sigma_{\rm v}$) and
spatial dispersion of trailing debris perpendicular to the Sgr plane ($\sigma_{Z_{Sgr,\odot}}$) 
are plotted as functions of present satellite mass $M_{\rm Sgr}$
for oblate ($q =$ 0.90, bottom row), spherical ($q =$ 1.0, middle row), and prolate ($q =$ 1.25, top row) models of the Galactic halo potential.
The solid lines represent the fiducial values found for 2MASS M giants from
Papers I and II, and the hatched areas show the regions that are within one standard deviation of these measurements.
Square points are for a series of simulations along a given orbit ($v_{\rm tan}=280/270/254$ km s$^{-1}$ for $q =$0.90/1.0/1.25 respectively) 
but with initial dwarf mass and scale length
pairs chosen to produce a similar central density.
Crosses represent simulations along these same orbits and in a similar initial mass range for the dwarf but with a variety of scale lengths, and
triangles represent simulations with fixed initial mass and scale length evolved along orbits with $v_{\rm tan}$ in the range
$\pm 20$ km s$^{-1}$ around 280/270/254 km s$^{-1}$.
\label{mplot}}
\end{figure}

Figure \ref{mplot} plots the calculated velocity dispersion (left-hand panels) and width (right-hand panels) as functions
of simulated bound satellite mass for choices of $q =$ 0.9 (lower panels), 1.0 (middle panels), and 1.25 (upper panels).  In all panels
the M giant dispersion/width is plotted as a solid line with 1-$\sigma$ error bars indicated by the hatched
regions, while the points in all panels indicate N-body simulation results (incorporating a 17\% artificial distance scatter
to simulate the photometric distance errors given in Paper I) for model satellites evolved along the orbits found earlier in \S 3.1.5C
for a variety of choices of initial satellite mass ($M_{\rm Sgr,0} = 10^7$ $M_{\odot}$ - $5 \times 10^9$ $M_{\odot}$) and 
physical scale ($r_{\rm 0} =$ 0.2 kpc - 1.5 kpc).

Clearly, similar values of $M_{\rm Sgr}$ are preferred for models in oblate, spherical,
and prolate Galactic potentials alike.  To quantify more precisely the range of acceptable masses indicated by Figure \ref{mplot}
we fit the data points in each panel with a third-order polynomial with 2.5-$\sigma$ rejection criteria iterated to convergence
and extrapolate from the resulting power-series coefficients the mass range whose $\sigma_{\rm v}$ and $\sigma_{\rm Z_{\rm Sgr,\odot}}$
lie within the 1-$\sigma$ uncertainty range around the M-giant measurements.
These results, presented in tabular form in Table \ref{mass.table}, indicate that in all models of the Galactic potential
considered the present bound mass of the Sgr dwarf should not be very different from
$M_{\rm Sgr} =$ 2 - 5 $\times 10^8 M_{\odot}$ if the model dwarf is to successfully reproduce the M giant observations.

\subsubsection{Constraining the Velocity of the Sgr Dwarf}

We now fix the initial mass and scale of the model dwarf such that the present-day dwarf has a bound mass in the range found
above in \S 3.2.1, and endeavor to refine our orbits using the single remaining free parameter $v_{\rm tan}$.  We explore a range of values
$\pm 20$ km s$^{-1}$ around the values $v_{\rm tan} =$ 280/270/254 km s$^{-1}$ chosen from test-particle orbits
previously in \S 3.1.5C.  Note that it is not possible to fix the final bound mass of the satellite in these simulations, since the change in the
orbital path produced by varying $v_{\rm tan}$ will naturally affect the mass-loss history of the model dwarf.  However, as demonstrated by
Figure \ref{mplot} (filled triangles) these small variations in $v_{\rm tan}$ have only a minor effect on the final mass of the model dwarf.

\begin{figure}
\plotone{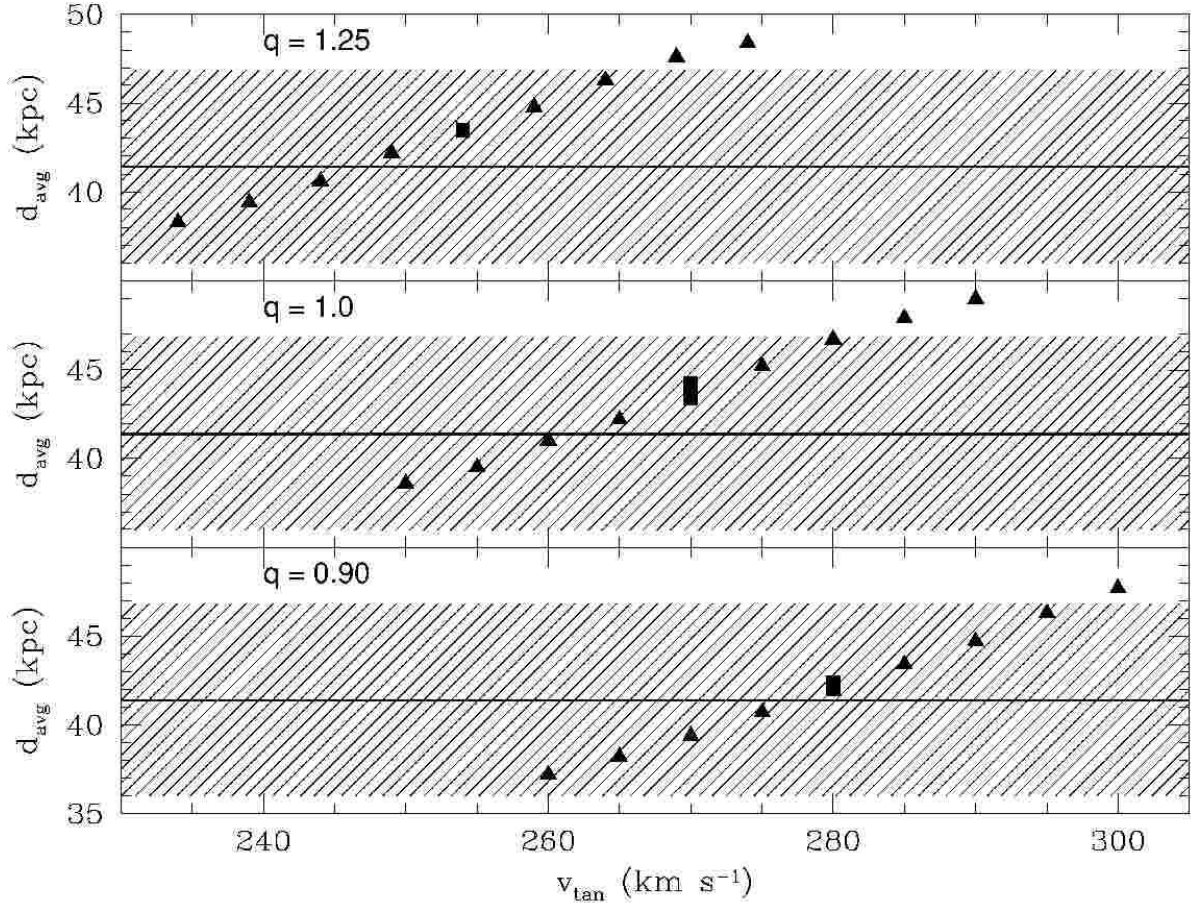}
\caption{Average distance of leading apoGalacticon debris ($d_{\rm avg}$),
plotted as a function of the tangential velocity parameter $v_{\rm tan}$ for oblate ($q =$ 0.90, bottom panel), 
spherical ($q =$ 1.0, middle panel), and prolate ($q =$ 1.25, top panel) models of the Galactic halo potential.
The solid lines represent the fiducial values found for 2MASS M giants from
Paper I, and the hatched areas show the regions that are within one standard deviation of those measurements.
Symbols are the same as in Figure \ref{mplot}, but only those square points and crosses which fall within the hatched regions on
Figure \ref{mplot} are included here.
\label{davgplot}}
\end{figure}

Returning to constraint 4 on the average apoGalacticon distance of leading debris,
the average distance of observed leading Sgr debris ($d_{\rm avg}$) is calculated from the 
2MASS database by averaging over the distances of
all stars in the range $\Lambda_{\odot} = 280^{\circ}$ - $320^{\circ}$ with heliocentric
distances $30$ kpc $< d_{\ast} <$ 60 kpc and subject
to the restrictions $E(B-V) < 0.555$, $1.0 < J-K < 1.1$, $|Z_{\rm Sgr,\odot}| < 5$ kpc, $Z_{\rm GC} > 10$ kpc
(this combination of restrictions was chosen to separate Sgr leading arm stars most clearly from the underlying disk population).
Figure  \ref{davgplot} plots the average apoGalacticon distance of the M giants 
as a solid line with 1-$\sigma$ error bars indicated by the hatched region, along with the values calculated from the simulated data (again incorporating 
a 17\% distance uncertainty) for the simulations with fixed initial mass and physical scale but varying $v_{\rm tan}$ (filled triangles).
Simulations with a range of initial masses and physical scales whose present bound mass falls within the acceptable range found in the previous section
are also plotted (filled squares and crosses): These points are difficult to distinguish since $M_{\rm Sgr}$ and $r_{\rm 0}$ are not the
primary factors governing the behavior of $d_{\rm avg}$, demonstrating the minor variation in $d_{\rm avg}$ permitted by the remaining uncertainty in satellite mass.
While Figure \ref{davgplot} shows a strong correlation between leading debris distance and orbital velocity however, the relatively large uncertainty in the
M giant debris distance allows us only to place constraints on the dwarf velocity to within about $\pm 20$ km s$^{-1}$.

\begin{figure}
\plotone{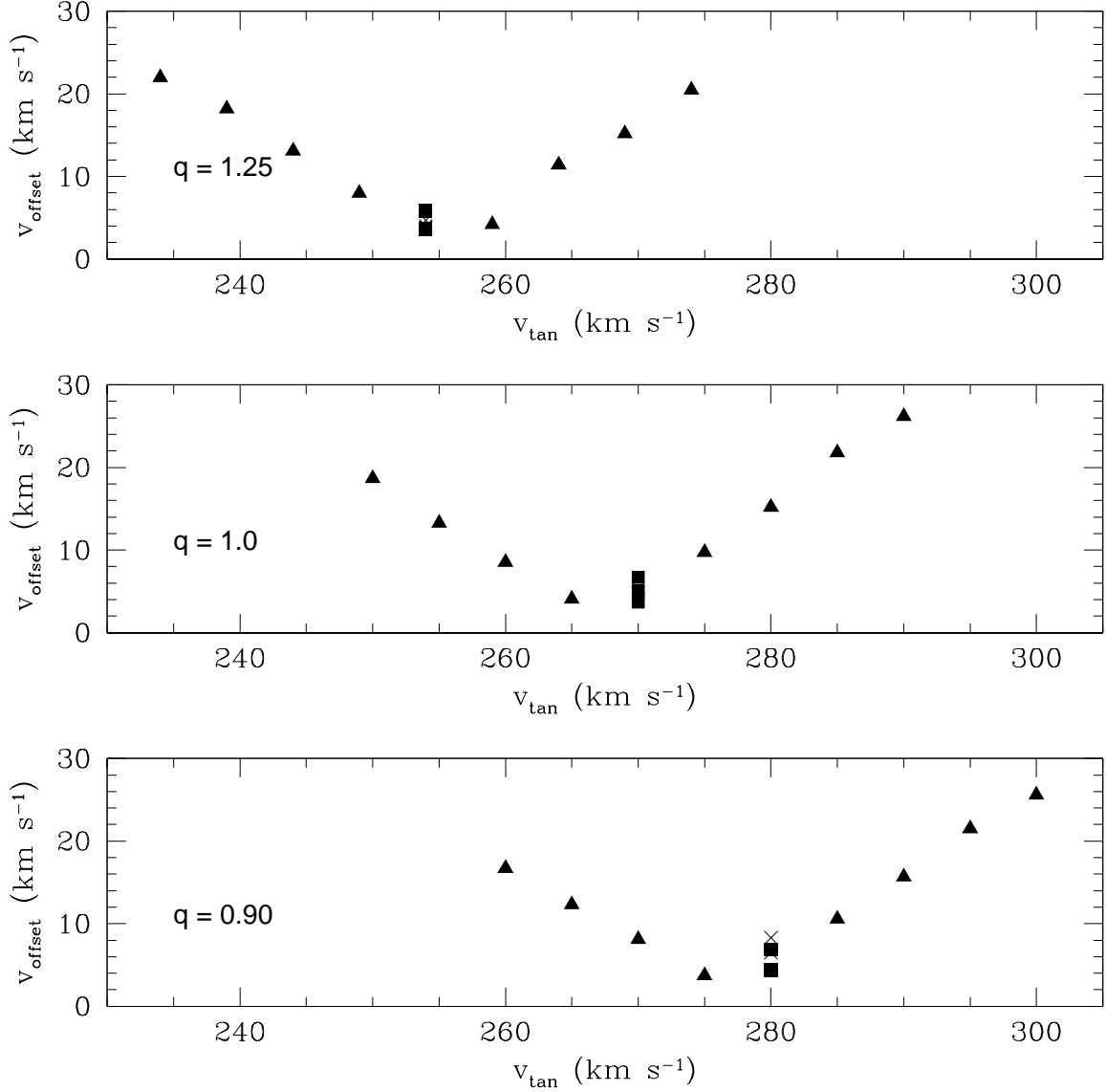}
\caption{Average offset of trailing debris velocities from the fiducial Sgr stream velocities (Paper II) for a range of choices
of the model Sgr dwarf velocity $v_{\rm tan}$ in
oblate ($q =$ 0.90, bottom row), spherical ($q =$ 1.0, middle row), and prolate ($q =$ 1.25, top row) models of the Galactic halo potential.
Symbols are the same as in Figure \ref{mplot}, but only those square points and crosses which fall within the hatched regions on
Figure \ref{mplot} are included here.
\label{voffsetplot}}
\end{figure}

A more compelling velocity constraint can be obtained by again using constraint 5, that the trailing arm velocities match those
observed for M giants.  We calculate the average offset of the centroid of simulated trailing debris velocities\footnote{In the interests of consistency
with previous analyses in Paper II, we again use the range $\Lambda_{\odot} = 25^{\circ}$ - $90^{\circ}$.} from the M giant centroid and plot
these offsets as a function of the tangential velocity of the dwarf in Figure \ref{voffsetplot}.  Well defined minima corresponding to the best
fits to the velocity data are obtained for specific velocities in each choice of the Galactic potential, and are fairly insensitive to the remaining
uncertainties in satellite mass (filled squares and crosses).  We therefore conclude that the best choices of tangential velocity for the model
dwarf are $v_{\rm tan} =$ 275-280/265-270/250-260 km s$^{-1}$ (note that, in this case, the best-fit test particle orbits obtained in \S 3.1.5C actually picked
out the best orbits for the N-body simulations).
Although each of these estimates are reasonably consistent with the observed value $v_{\rm tan} \approx v_{\rm b} =$ 280 $\pm$ 20 km s$^{-1}$ measured
by Ibata et al. (2001), it is interesting to note that the Ibata et al. (2001) measurement appears to slightly favor oblate models of the Galactic
halo over prolate models at the 1-$\sigma$ level for our current choice of $v_{\rm circ, \odot} =$ 220 km s$^{-1}$.  Note, however, that a higher value
of $v_{\rm circ, \odot}$ will systematically shift these estimates of $v_{\rm tan}$ to higher velocities (see Figs. 4 \& 5, dashed line in lower right-hand panels), 
resulting in better agreement of estimates of 
$v_{\rm tan}$ in prolate halos with the Ibata et al. (2001) measurement.


\subsubsection{Our best-fit model}
Based upon Figures  \ref{mplot}, \ref{davgplot}, and \ref{voffsetplot}, simulations with 
$M_{\rm Sgr} =$ 2.6-5.0/2.5-5.3/2.5-5.5 $\times 10^8 M_{\odot}$
and $v_{\rm tan} =$ 275-280/265-270/255-260 km s$^{-1}$ best fit our constraints, and these models are hereafter referred to
as our ``best-fit models''\footnote{Complete 
data files of model Sgr debris from these best-fit models are provided on the web at
http://www.astro.virginia.edu/{$\sim$}srm4n/Sgr/ to aid future comparisons of these models with new observations and new
disruption models.}.  
Although the uncertainty in the Galactic potential gives rise to uncertainties in $v_{\rm tan}$ considerably greater than the ranges given here,
within a given potential $v_{\rm tan}$ can be constrained to within about $\pm$ 5 km s$^{-1}$.
Our best-fit models have a maximum extent of bound material  $r_{\rm bound}\sim 500'$ along the semi-major axis, within which
we calculate a luminosity for Sgr of $L_{\rm Sgr} = 1.4 \times 10^7 L_{\odot}$ using data presented in Paper I.
The mass-to-light ratio of Sgr in these models should therefore
be $M_{\rm Sgr} / L_{\rm Sgr} =$ 19-36/18-38/18-39 $M_{\odot} / L_{\odot}$.  
While the 500' maximum extent for bound material is somewhat dependent on the adopted internal
structure of the satellite, it is on the order of the true tidal radius previously pointed out (\S4.3.3
of Paper I) as required to avoid Sgr having a quite extraordinary (and unlikely) bound mass, and is also of
order the observed {\it minor} axis dimension (i.e., 0.35 times that of the 1801' major axis radius) 
of the limiting radius of the fitted King profile to the central satellite.

These orbits have periods of 0.85/0.88/0.87 Gyr
with periGalactica and apoGalactica of 10-16/14/14-19 kpc and 56-58/59/56-59 kpc respectively\footnote{Note that {\it ranges}
are given for non-spherical potentials since for such non-spherically symmetric potentials the apoGalacticon and periGalacticon distances are dependant
upon the polar angle of the satellite, and hence these distances may vary slightly from orbit to orbit.},
and a present space velocity $(U,V,W)$\footnote{We adopt a right-handed Galactic Cartesian coordinate system
with origin at the Galactic Center.} $=$ (238, -42, 222)/(235, -40, 213)/(231, -37, 198) km s$^{-1}$,
corresponding to $(\Pi,\Theta,Z) = $(230, 75, 222)/(227, 73, 213)/(224, 69, 197) km s$^{-1}$ and 
$(v_{\rm r}, v_{\rm b}, v_{\rm l}) =$ (171, 272, -65)/(171, 263, -63)/(171, 247, -59) km s$^{-1}$ with respect to the Galactic standard of rest.
These velocities will scale roughly with the assumed value of $v_{\rm circ, \odot}$, although will also depend systematically upon $q$, $d$, and $\alpha$.

\begin{figure}
\plotone{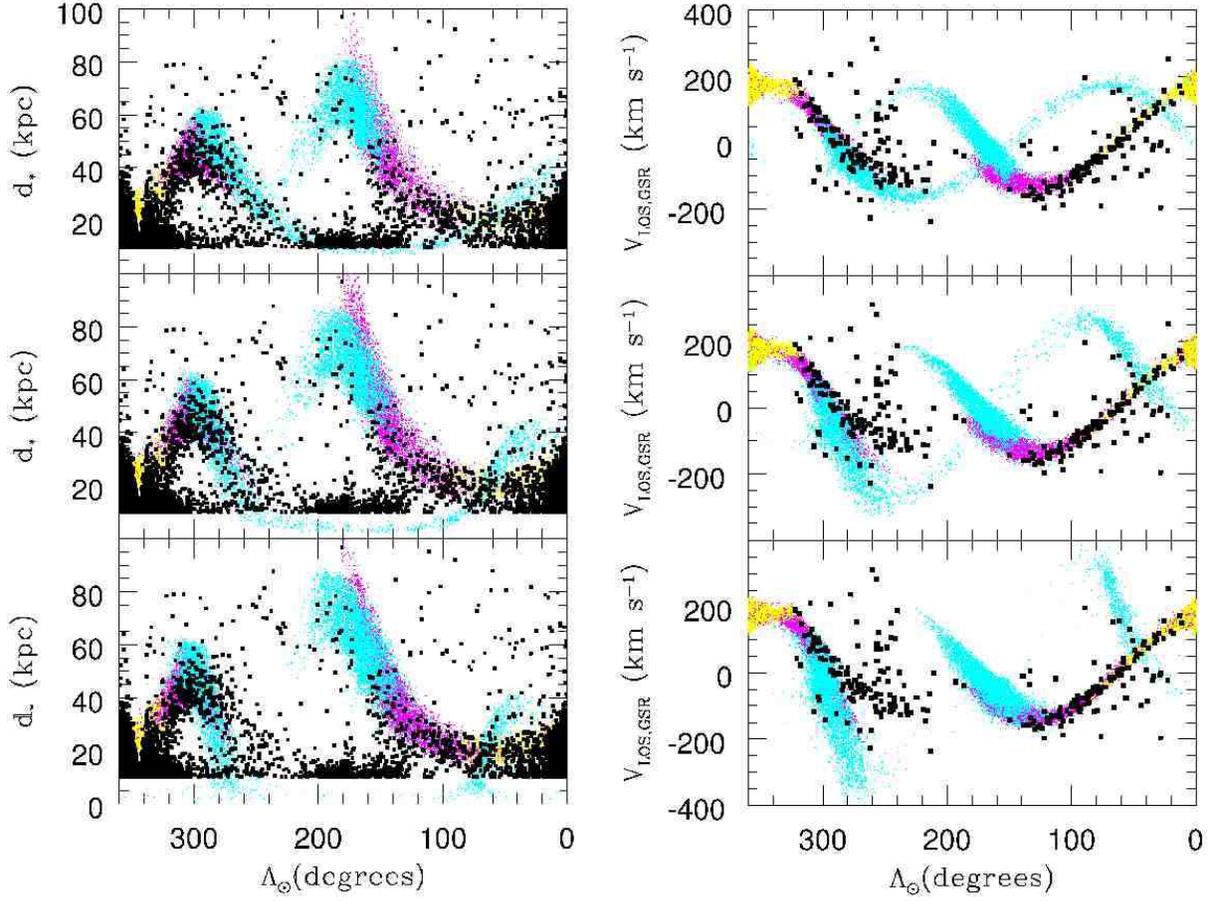}
\caption{Distance and velocity data are plotted as a function of orbital longitude for simulated satellite debris
from the best-fit models in oblate ($q =$ 0.90, bottom row), spherical ($q =$ 1.0, middle row), and 
prolate ($q =$ 1.25, top row) models of the Galactic halo potential (colored points)
and 2MASS M giant data from Papers I, II, and V (black points and solid squares, compare to Figs. 10 and 6
of Papers I and II respectively).
A 17\% artificial random distance scatter has been applied to simulated debris particles
to mimic the photometric distance error present in the 2MASS sample.  Note that M giants closer than 10 kpc have been omitted from
the lower panel in order to show nearby simulated debris.\label{simplot1}}
\end{figure}

Figure \ref{simplot1} plots simulated Sgr debris for our best-fit models along with the M giant distance and
velocity data from Papers I, II and V, and demonstrates visually that our models generally fit the M giant observations well.
The M giant data is clearly traced by debris released during the last two pericentric passages of the model dwarf
(yellow and magenta points) and possibly by debris released three pericentric passage ago (cyan points), although there appear to be
far fewer M giants corresponding to cyan points than magenta or yellow.  This corresponds to M giants becoming unbound from the Sgr dwarf over the last 1.5 - 2.5
Gyr --- consistent with constraint 10, that the debris age be younger than the typical age of an M giant star.
Note, however, that as predicted by the orbits in \S 3.1 models in oblate and spherical halo potentials fail to fit the leading velocity trend
(particularly for cyan points), while
the model orbiting in a prolate potential both reproduces this velocity trend and provides a more convincing fit to the apparent trend of M giant
distances at $\Lambda_{\odot} \sim 220^{\circ}$ - $260^{\circ}$.  Note also the presence of cyan and green debris within a few kpc of the Sun
over a wide range of $\Lambda_{\odot}$ for oblate and spherical halo models - this is a consequence of the leading streamer diving almost
directly through the Solar Neighborhood in these two models.  Conclusive proof of the presence or absence of Sgr debris around the Sun
would provide a significant additional constraint on the models.

\begin{figure}
\plotone{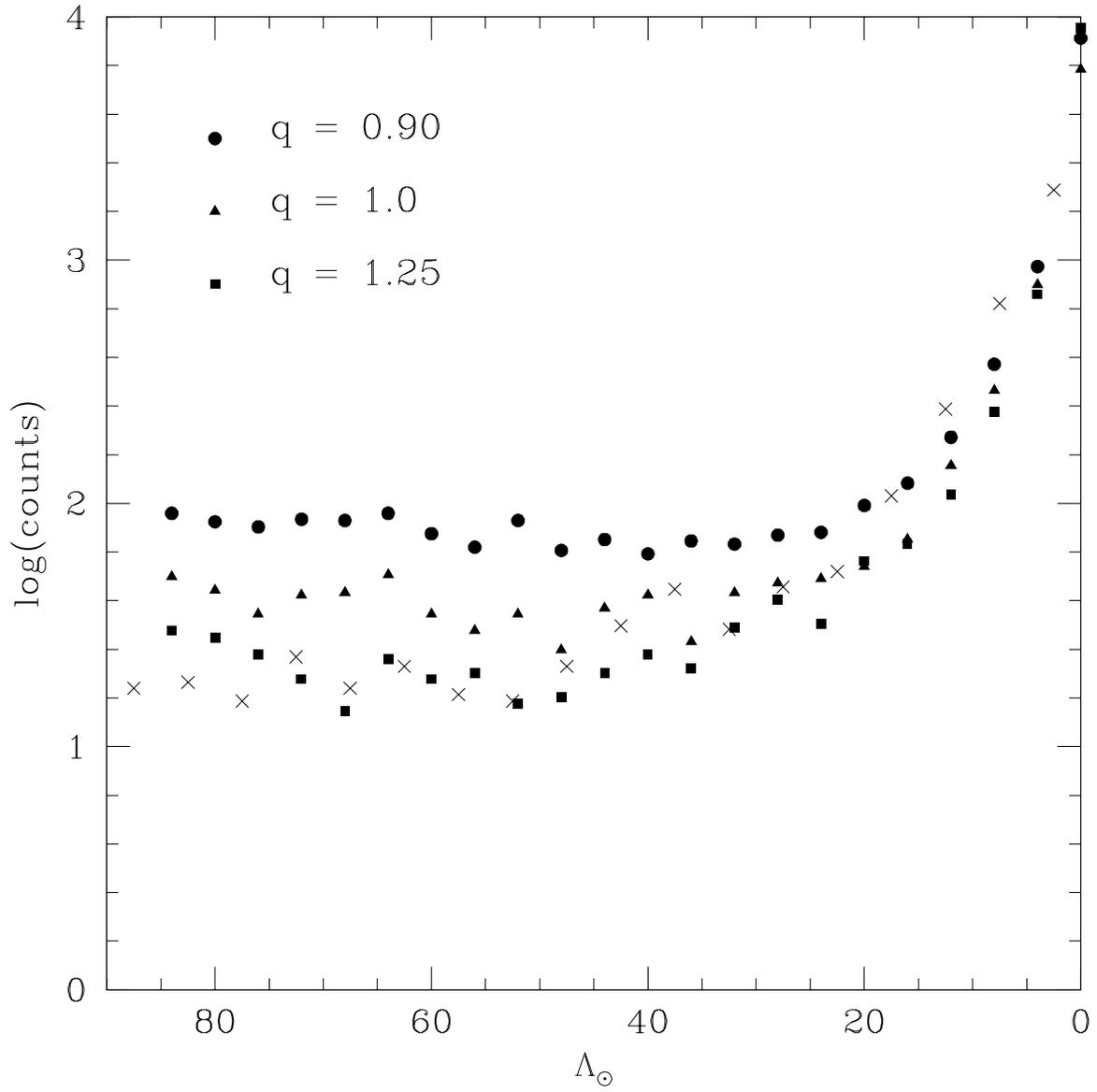}
\caption{Counts (per $4^{\circ}$ of orbital longitude) of debris along the trailing tail,
data are shown for the best-fit models (filled circles/triangles/squares) and for background-subtracted M giant data (crosses) from
Paper I.\label{denplot}}
\end{figure}

The density of stars in the trailing stream for the best-fit models is plotted as a function of $\Lambda_{\odot}$
in Figure \ref{denplot}, and is similar in structure to the density of the M giant stream (constraint 11, plotted in Fig. 13
of Paper I), with a break in the slope of the density profile around $\Lambda_\odot=20$ degrees and a relatively 
constant density thereafter (we only consider this first break in the observed
density profile since we expect this to depend primarily
upon satellite mass).  The details of the run of density along the trailing streamer will depend on the internal light
distribution of the satellite.  However, since we consider only single-component models in this paper, we omit further consideration of the density profile
and internal structure of the dwarf at this time.

\section{DISCUSSION}
\subsection{Comparisons with Previous Data}

As noted in \S 2.2, most other Sgr detections around the sky fall within
the M giant-traced tails (see Fig. 17 of Paper I), so that our best-fit models also provide a good match to these other data. In 
this section, we compare our predictions for older Sgr debris (green points) not traced by the M giants with observations of older tracers.

\begin{figure}
\plotone{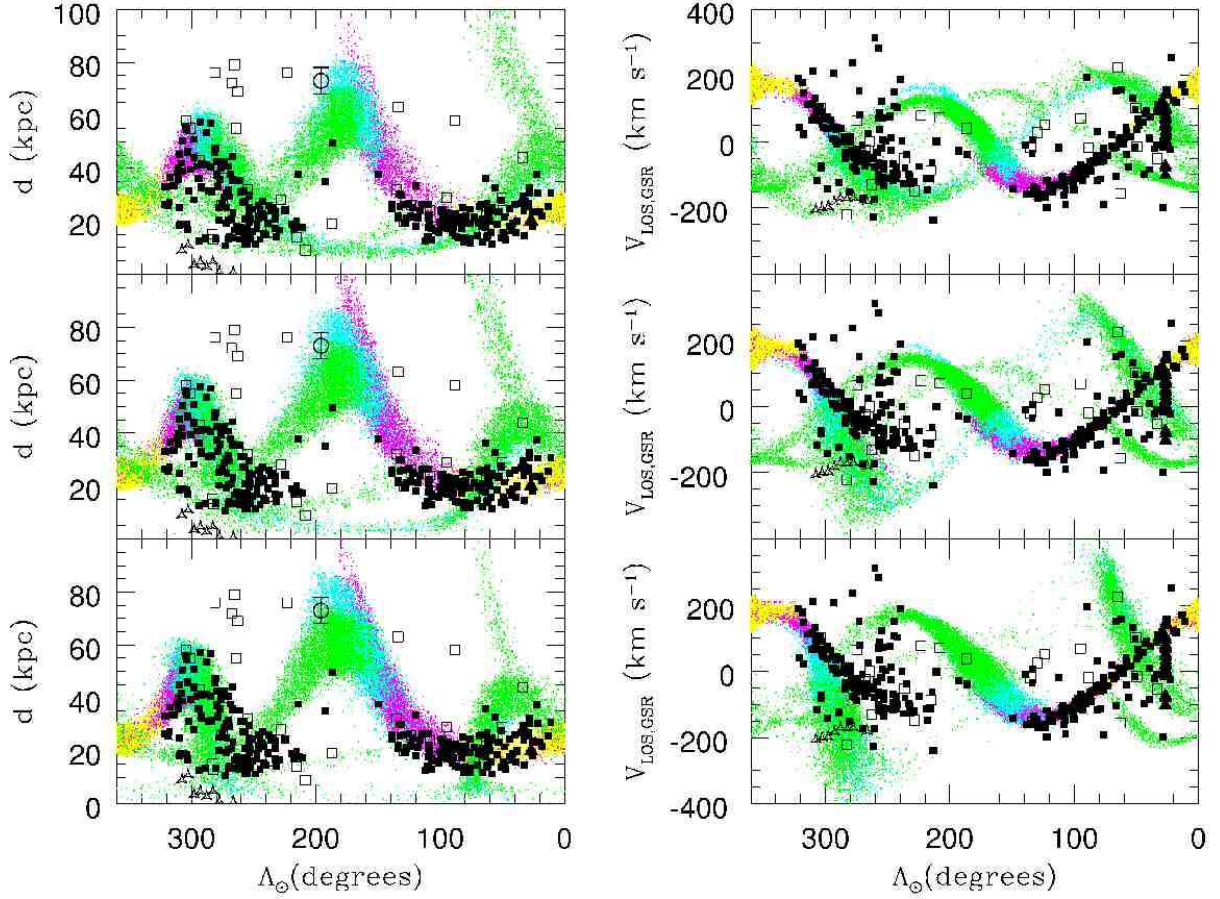}
\caption{Distances and radial velocities of debris from the best-fit
models in oblate ($q =$ 0.90, bottom row), spherical ($q =$ 1.0, middle row), and 
prolate ($q =$ 1.25, top row) models of the Galactic halo potential (colored points)
are overplotted with data from selected recent observations.
Filled squares denote data from Papers II and V, open boxes represent carbon stars selected from Totten \& Irwin (1998), solid 
triangles are data from Majewski et al. (1999), open trianges are data from Kundu et al. (2002), and the open circle is
from Newberg et al. (2003).
A 17\% artificial random distance scatter has been applied to simulated debris particles
to mimic the photometric distance error present in the 2MASS sample.
\label{simplot2}}
\end{figure}

In Figure  \ref{simplot2}, carbon star data\footnote{Carbon stars have been selected from 
Totten \& Irwin (1998) subject to the requirement that both distance 
and velocity data have been measured, and also
subject to the photometric criteria employed by Ibata et al. (2001) that 11 $< R < 17$ and $B_J - R > 2.5$.} 
(open boxes)
are plotted for comparison with our best-fit Sgr models (colored points).
While some of the carbon stars appear consistent with both M giant and simulated debris, many
others have distances and velocities that differ substantially from the M giant and model distributions, and attempts
to fit simulation models to these carbon stars will likely produce results that differ noticeably from our own best-fit models
and the M giant data.
Although some of this discrepancy could be due to the uncertain distance scale for the carbon stars (see \S 8.3
of Paper I), it is also
possible that these stars could trace debris 
older than the $\sim$ 2.5 Gyr old M giant stream, 
since carbon stars
can have 
larger ages (5 - 6 Gyr) than M giants.

The open triangles near $\Lambda_{\odot} \sim 300^{\circ}$ in Figure  \ref{simplot2} 
represent data for a set of metal-poor, K-giant stars first pointed out by \cite{kundu02}. 
Using semi-analytical modeling, \cite{kundu02} proposed that these stars represent debris stripped from Sgr 
three pericentric passages ago (corresponding to cyan-colored points in our model). 
Indeed, our model suggests that these points may plausibly be fit by cyan or green debris (i.e. debris from 3 - 4 pericentric
passages ago) in the $q =$ 0.90 leading streamer that is 
currently raining down from the North Galactic Pole onto the Solar Neighborhood, although the interpretation of these data
is uncertain in models where $q =$ 1.0 or 1.25.

We also note an interesting comparison with possible Sgr red clump stars detected in a pencil-beam survey by Majewski et al. (1999)
at $(l,b) = (11^{\circ},-40^{\circ})$, and for which the
radial velocity data are plotted in Figure \ref{simplot2} (top panel, solid triangles).
These stars at $\Lambda_{\odot} = 27^{\circ}$ exhibit a range of line-of-sight velocities from 0 to 150 km s$^{-1}$,
which closely matches the predicted range of velocities of simulated 
leading tidal debris wrapped almost $360^{\circ}$ in orbital longitude from 
the Sgr dwarf (cyan and green points) for simulations where $q =$ 1.0 or 1.25.  The degree of this agreement is highly sensitive to the mass of the model
satellite: Simulations with present mass $M_{\rm Sgr} = 5 \times 10^8 M_{\odot}$ 
predict a larger dispersion in velocities than observed by Majewski et al., while
simulations with mass $M_{\rm Sgr} = 2 \times 10^8 M_{\odot}$ 
predict a smaller dispersion than observed.  It is tempting therefore to point to these data
as further evidence in favor of the satellite mass estimates determined earlier in \S 3.2.1
However, the distance to these stars is measured to be roughly 20 kpc (Majewski et al. 1999) --- about 
half that of the cyan - green leading debris whose velocities they reproduce so well --- and therefore, while they are interesting
to compare to model data, their true origin and interpretation remains unclear.


Recently, the discovery of an overdensity of A-colored stars in the Sloan Digital Sky 
Survey with apparent magnitude $g_0\sim 20.3$ at $\Lambda_\odot=187^{\circ}-212^{\circ}$ degrees 
and within 15 kpc of Sgr's nominal orbital plane was 
announced (Newberg et al. 2003). These authors estimate an average heliocentric distance of 83 kpc
to these stars, but note that other detections 
in directions which overlap the M giant stream suggest that their adopted distance scale is 12.5\% larger than that
used to calibrate the M giants in Paper I. The open circle in Figure \ref{simplot2} (left-hand panels) plot the average of their data, 
with the distance rescaled to 73 kpc so that the 
M giant and SDSS distance scales
match. 
Figure  \ref{simplot2} suggests that it is plausible 
to identify the SDSS detection with debris of age $\sim$ 1.5 - 2.5 Gyr (i.e. cyan-colored points) in the trailing Sgr stream, although
future radial velocity measurements could help determine whether this identification is correct or if the 
Newberg et al. feature
is instead
a part of some older, more distant section of the stream or even halo substructure unrelated to Sgr.
Newberg et al. (2003)
also note a hint of precession in the Sgr stream by comparing their detections of leading and trailing debris closer to Sgr's core, 
in agreement with our own results presented in Paper III.
Unfortunately, the angular extent of the 83 kpc debris has not yet been mapped accurately 
enough to pinpoint the angular position of the centroid of the debris; such a measurement could in the future provide a strong
constraint on the flattening of the Galactic potential.

\subsection{Comparisons with Previous Sgr Simulations}
Previous attempts to model the orbit and disruption history of the Sgr dwarf 
(e.g. Velazquez \& White 1995, Johnston, Hernquist \& Bolte 1996, 
Ibata et al. 1997, Edelsohn \& Elmegreen 1997, Ibata \& Lewis 1998, G{\' o}mez-Flechoso, Fux \& Martinet 1999,
Johnston et al. 1999, Helmi \& White 2001, Ibata et al. 2001, Mart{\' i}nez-Delgado et al. 2004)
have made considerable progress in constraining models of the dwarf using only the previously available
pencil-beam detections of satellite debris.
In this paper we have presented the first model based upon a complete all-sky view of the satellite's tidal streams,
and in this section we review and compare some of the predictions of these earlier models to those of our own best-fit models.

We first consider those results for which the majority of simulations by different groups have generally converged.
Almost all simulations agree that the radial period of the Sgr dwarf should be about 3/4 Gyr:
In this work we find a period for our best-fit models of 0.85/0.88/0.87 Gyr, in reasonable agreement with previous estimates of 
0.76 Gyr (Velazquez \& White 1995, Ibata et al. 1997), 0.7 Gyr (Ibata \& Lewis 1998), 0.55-0.75 Gyr (Johnston et al. 1999),
0.85 Gyr (Helmi \& White 2001), and 0.74 Gyr (Mart{\' i}nez-Delgado et al. 2004).
There is a little more spread in the estimates proposed by different groups for the
periGalacticon and apoGalacticon distances of the dwarf's orbit:  Previous estimates
include (respectively) 10 kpc and 52 kpc (Velazquez \& White 1995), 15 kpc and 60 kpc (Ibata \& Lewis 1998),
15 kpc and 70 kpc (G{\' o}mez-Flechoso, Fux \& Martinet 1999), 13 kpc and 41 kpc (Johnston et al. 1999),
16 kpc and 60 kpc (Ibata et al. 2001), and 12 kpc and 60 kpc (Mart{\' i}nez-Delgado et al. 2004).
With the 2MASS database it is possible to measure the apoGalacticon of leading tidal debris directly, and we match
this constraint best by using models for Sgr that have orbits with periGalacticon and apoGalacticon distances of 
10-16/14/14-19 kpc and 56-58/59/56-59 kpc respectively.  We note, however, that the distance scale assumed for the M giants in Paper I
is not yet secure,
and that the estimated size of Sgr's orbit may scale accordingly.

Among those areas in which common values among the disruption models presented by various groups have not yet been found,
perhaps foremost is the $V$ component of the Galactic $(U,V,W)$ velocity of the Sgr dwarf.
Some simulations (e.g., Ibata et al. 1997) have simply set $V=0$ km s$^{-1}$ (thereby assuming a polar orbit)
since this component was so poorly
known.  Now that we have an accurate measurement of Sgr's orbital pole (Paper I), we are able to predict the direction of its motion more precisely. 
Based on our best-fit model, we predict that the proper motion of the Sgr dwarf should be
$\mu_{\rm l} \cos(b) = -2.59/-2.57/-2.54$ mas yr$^{-1}$ and $\mu_{\rm b} = 2.26/2.18/2.05$ mas yr$^{-1}$ in the Solar rest frame\footnote{
We adopt a solar peculiar velocity of $(U,V,W) = (9,12,7)$
km s$^{-1}$ relative to the LSR, for which we adopt a rotation velocity of 220 km s$^{-1}$ (\S 3.1.5B).}.
The direction of this proper motion prediction ($\mu_{\rm b}/\mu_{\rm l} \cos(b)=0.87/0.85/0.81$) is expected to be fairly robust within
potentials with each choice of $q$.
However, the amplitude of the proper motion will depend on the exact form of the Galactic potential, and hence should be revised
once other fundamental Galactic parameters such as $R_\odot$ and $v_{\rm circ, \odot}$ are known more precisely. 
Conversely, as more accurate measurements of Sgr's proper motion become available 
it will be possible to refine constraints on the Galactic rotation curve.

A second area of debate concerns the present bound mass of the Sgr dwarf, for which estimates range from 
$M_{\rm Sgr} = 7.0 \times 10^6 M_{\odot}$ (Mart{\' i}nez-Delgado et al. 2004) 
to $M_{\rm Sgr} = 1.0 \times 10^9 M_{\odot}$ (Ibata et al. 1997).
Helmi \& White (2001) find an intermediate value for a purely stellar satellite model with initial mass
$M_{\rm Sgr,0} \approx 5.0 \times 10^8 M_{\odot}$.
As demonstrated in \S 3.2.1, we find that a range of final masses $M_{\rm Sgr} = 2$ - $5 \times 10^8 M_{\odot}$
yield
tidal tails whose thickness and velocity dispersion are consistent with M giant measurements
in oblate, spherical, and prolate models of the Galactic potential.
Using Figure  \ref{mplot} we conclusively rule out models with a mass far outside this range (such
as that of Mart{\' i}nez-Delgado et al. 2004), 
since models with very high or low masses will not be able
to produce tidal tails with the observed thickness and dispersion.  
Visual inspection of the figures in 
Mart{\' i}nez-Delgado et al. (2004) appears to contradict this statement.
However, these authors' simulation embeds the model satellite in a 40,000 particle 
live halo, which is probably responsible for the width of the 
debris stream: Earlier work (Johnston, Spergel \& Haydn 2002)
has found that significant heating of a Sgr-like debris stream can occur in a simulation using a live halo, 
even in a halo model realized with $10^6$ particles.

As another consequence of the smaller satellite mass used in their model, Mart{\' i}nez-Delgado et al. (2004) predict leading debris at 
$(\alpha ,\delta) = (210^{\circ}, 0^{\circ})$ (corresponding to
$\Lambda_{\odot} = 284^{\circ}$) to be composed of stars which have been unbound 
from the satellite for 5 Gyr or more (since debris from lower-mass satellites takes longer to spread along the
orbit), in contrast to the roughly 2 Gyr found by our own analysis.
As demonstrated in Figure  \ref{simplot1}, the Sgr M giants --- 
which have an estimated age of 2-3 Gyr  --- are visible to at least this point in the leading tidal stream.
As Mart{\' i}nez-Delgado et al. (2004) point out (and we discuss in Paper I), 
stellar populations formed in the densest central regions of the satellite
should not be immediately reflected in the tidal streams, and it will take some time for these stars to be present in any quantity 
in the outer regions of the satellite.  
Hence, it is unlikely that the M giant population became mixed into the outer
regions of Sgr within a small fraction of a Gyr, and
we consider the mean age estimate of 5 Gyr for this section of the tidal stream to be too high.

\subsection{Evolution of Sgr's Orbit}
In Paper III we showed that only Galactic potentials with oblate halos could reproduce the
precession of the orbital plane apparent in the leading {\it vs} trailing data sets.
In contrast, \citet{helmi04} demonstrated that only Galactic potentials with
prolate halos could reproduce the
velocity trends in the leading debris.
In this paper (\S3.1) we explore a much wider variety of Galactic
potentials than has been considered previously but fail to find a
single orbit that can fit both the velocity trends and sense of precession.
Our conclusion is that the  assumption of non-evolution of the orbit over the
time-period that the debris explores is incorrect. 

\begin{figure}
\plotone{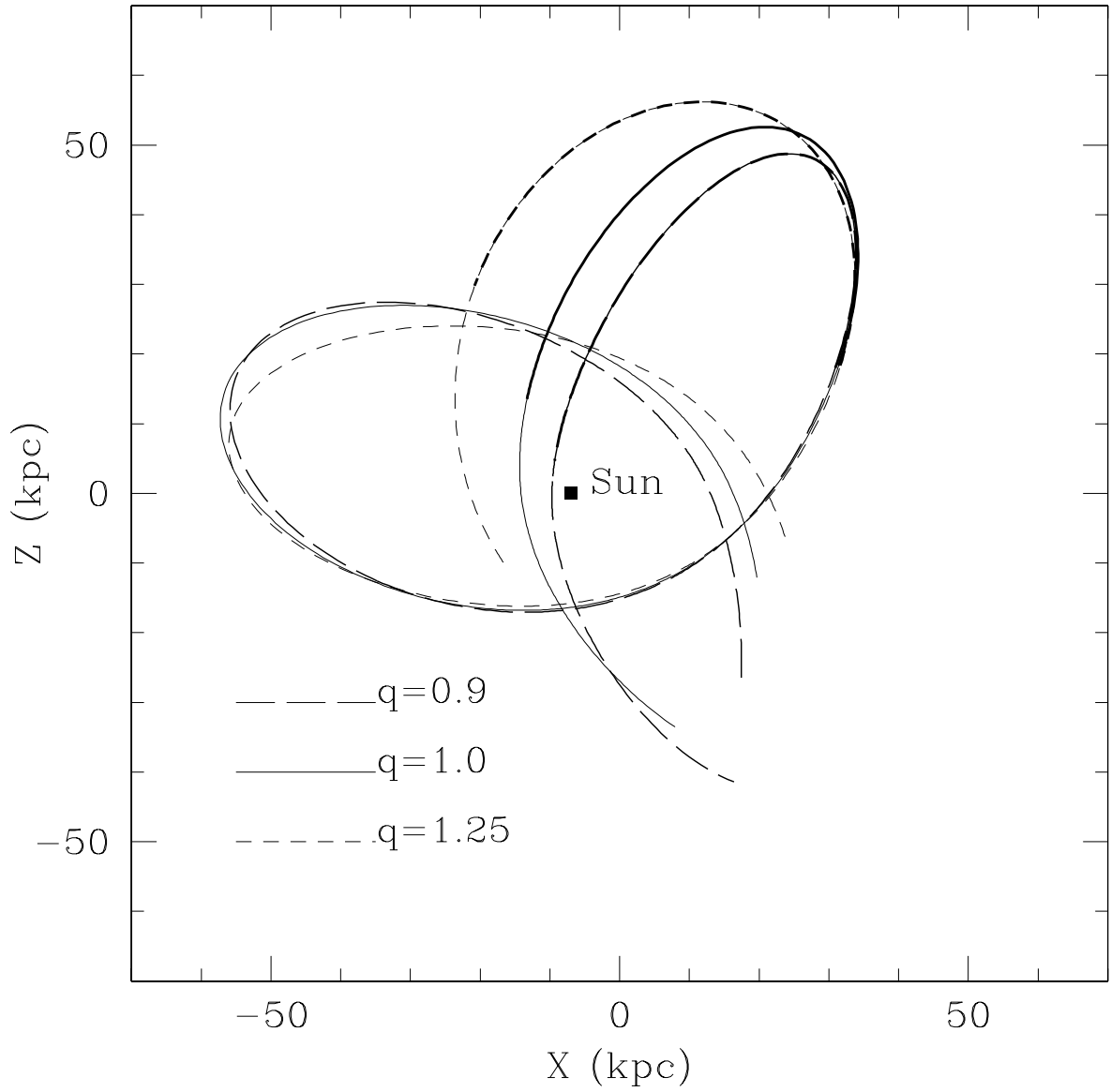}
\caption{Plots in Galactic coordinates of the same orbits shown in Figure \ref{orb.fig}.}
\label{orbxz.fig}
\end{figure}

Since simulated debris in the region with troublesome velocities 
is cyan and green (lost 2 and 3 orbits ago respectively), 
we estimate the timescale over which the
evolution has taken place to be $\sim 2-3$ Gyrs.
We can get some idea of the physical scale of the evolution necessary by looking at the difference
between the orbits in prolate, spherical and oblate potentials that is responsible
for the difference in the velocity trend.
Figure \ref{orbxz.fig} plots the orbits shown in Figure \ref{orb.fig} in Galactic
coordinates with the region corresponding to the leading debris velocity data
shown as bold along each curve.

As the potential moves from prolate to oblate, the orbit passes progressively nearer the
Sun and line-of-sight velocities more closely reflect the full motion along the orbit.
This explains why the simulated line of sight velocities in this region become more extreme
with the oblateness of the potential.
Figure \ref{orbxz.fig}
also suggests that observed debris velocities in the leading region might be 
accounted for
even in an oblate or spherical potential if the pericenter of Sgr's orbit 
has decreased by a factor of order unity within the last 2-3 Gyrs (from visual
inspection of the figure) since such a decrease in pericenter of the Sgr {\it dwarf} over time
could shift older Sgr {\it debris} out to greater distances from the Sun corresponding to the greater pericenter
of the dwarf on the passage on which the debris became unbound.
Three factors could contribute to this evolution:
\begin{description}
\item{An encounter with a large lump in the Milky Way potential},
 either dark or luminous \citep[e.g. such as the Large Magellanic
 Cloud, see][for a full description of this idea]{zhao98}: We consider
 this unlikley since we would expect the signature of such an event to be
a sudden change in Sgr's orbit,
 and a corresponding sudden change in the velocities along its debris, rather than the
 smooth trends seen.
\item{Global evolution of the Galactic potential:} We also consider this unlikely
since: (i) The evolution would have to be very large in order to bring the
pericenter inwards by a factor of two in such a short amount of time; and (ii) 
any global evolution would affect both Sgr's and the debris' orbits similarly.
\item{Dynamical friction:} 
If we re-arrange equation [7-27] from \citet{binney87} we can find the mass necessary
$M_{\rm fric}$ for a circular orbit at $r=30$ kpc (i.e.
to represent an orbit with of order unity larger pericenter than Sgr today)
to decay to the center of the
Galaxy over a time period
$t_{\rm fric}=2$ Gyrs in a Galaxy with a flat rotation curve and $v_{\rm circ,\odot}=220$ km s$^{-1}$:
\begin{equation}
  M_{\rm fric}={1 \over \ln \Lambda} \left({1.0\times 10^{10} {\rm Gyrs} \over t_{\rm fric}}\right)
\left({r \over 60 {\rm kpc}}\right)^2
\left({v_{\rm circ,\odot}  \over 220 {\rm km\,s^{-1}}}\right) 2\times10^{10} M_\odot.
\end{equation}
\citet{binney87} estimate $\ln \Lambda \sim 3$ for the combined Large
and Small Magellanic Clouds.  Since we expect $\Lambda \propto
1/M_{\rm fric}$, and know the current mass of Sgr to be 2 - 5 $\times 10^8 M_{\odot}$, we expect
$\ln \Lambda=5-9$ to be the relevant range for our own estimate and
hence $M_{\rm fric}\sim 2.5-5 \times 10^9 M_\odot$.  Moreover, we
consider this only a lower limit on the necessary mass since Sgr's
orbit is not circular.  \citep[See][for a general discussion of
dynamical friction acting on Sgr over a Hubble time.]{jiang00,zhao04} 
\end{description}


Although dynamical friction
seems like the most favourable explanation for the orbit evolution it
does require Sgr to be an order of magnitude more massive just 2 Gyrs
ago and debris lost at that time in our mass-follows-light models
would have a correspondingly larger
dispersion in velocity (by a factor of order $\sqrt{10}=3$) and
distances.  Since the observed velocity dispersion in the debris in the
discrepant, leading portion of the stream is actually quite similar to
that seen in our simulations ($\sim 17$ km s$^{-1}$, see Fig. \ref{simplot1}) this suggests that, in order to fit the data,
in addition
to dropping our assumption of a single orbit for Sgr,
we will also have to
move beyond modelling Sgr as a {\it single component} system. 
Hence,  while the mean trend in the leading streamer will tell
us how much total mass needs to have been lost from Sgr, 
the low dispersion offers the additional opportunity of constraining
how much more tightly 
bound the light matter is compared to the dark matter.
A study of these combined effects is in progress.

\section{SUMMARY}

In this paper we have presented the first model of the tidal tails of the 
Sgr dwarf galaxy based upon a coherent, all-sky picture of the system 
in both position and radial velocity, as represented by M giants selected from the 2MASS database. 
We summarize our conclusions as follows:
\begin{description}
   \item{Shape and evolution of the Galactic potential ---}  In a companion paper (Paper III) we have shown that {\it oblate} ($q =$ 0.90) models of
the Galactic halo potential best reproduce the difference in orbital poles between leading
and trailing M giant tidal debris, while in this paper (see also \citet{helmi04}) we find that {\it prolate} ($q =$ 1.25) models are
required to reproduce the trend of observed M giant leading debris velocities.  Although we explore a wide variety of Galactic potentials
we fail to find a single orbit that can simultaneously reproduce the observed orbital pole precession and leading debris velocity trend,
and conclude that some evolution of the orbit of Sgr has occured over the past few Gyr.
   \item{Mass of the Milky Way Galaxy ---}  Within our simulations that best reproduce the observed Sgr dwarf tidal tails, the enclosed
mass of the Milky Way within 50 kpc is found to be $3.8-5.6 \times 10^{11}M_\odot$.
   \item{Mass of the Sgr dwarf ---}  The present bound mass of the Sgr dwarf has been restricted to the range
$M_{\rm Sgr} = 2$ - $5 \times 10^8 M_{\odot}$, constrained by the width and velocity dispersion of the trailing M giant
tidal tail.  Taking $L_{\rm Sgr} = 1.4 \times 10^7 L_{\odot}$ as the luminosity of Sgr,
this gives a range of possible values for the mass-to-light
ratio of Sgr from $M_{\rm Sgr} / L_{\rm Sgr} = 14$ - 36. Although all of our models maintained cores of bound material, we would expect similar dispersions to 
be seen in the debris if a similar amount of recently
unbound (i.e. on the current pericentric passage) mass were present within the same distance scale.
   \item{Orbit of Sgr ---}  The Sgr orbit in our best-fit models ($q =$ 0.90/1.0/1.25) has a pericenter of  10-16/14/14-19 kpc, 
an apocenter of 56-58/59/56-59 kpc and a radial time period of 0.85/0.88/0.87 Gyr. 
These values depend on the distance scale adopted for the M giants and the exact form of the Galactic potential.  
   \item{Proper motion of Sgr ---}  For our best-fit models ($q =$ 0.90/1.0/1.25), the tidal tails of the dwarf as traced 
by 2MASS M giants are best reproduced
by a satellite situated at $(X,Y,Z) =$ (16.2, 2.3, -5.9) kpc with velocity tangential to the line-of-sight
$v_{\rm tan} =$ 275-280/265-270/255-260 km s$^{-1}$ corresponding to
space velocities $(U,V,W) =$ (238, -42, 222)/(235, -40, 213)/(231, -37, 198) km s$^{-1}$,
i.e. $(\Pi,\Theta,Z) = $(230, 75, 222)/(227, 73, 213)/(224, 69, 197) km s$^{-1}$,
$(v_{\rm r}, v_{\rm b}, v_{\rm l}) =$ (171, 272, -65)/(171, 263, -63)/(171, 247, -59) km s$^{-1}$, and proper motion
$\mu_{\rm l} \cos(b) = -2.59/-2.57/-2.54$ mas yr$^{-1}$ and $\mu_{\rm b} = 2.26/2.18/2.05$ mas yr$^{-1}$ in the Solar rest frame.
These velocities are dependent on the model assumed for the Galactic potential, and will scale roughly with choice of $v_{\rm circ,\odot}$
and other Galactic parameters.
   \item{Solar neighborhood debris ---}  Our best-fit models orbiting in oblate ($q = 0.90$) and spherical ($q = 1.0$) potentials
predict that the Sun is currently bathing in a stream of 
debris from Sgr, passing both inside and outside the Solar Circle.  However, models orbiting in prolate ($q =$ 1.25) potentials
are inconsistent with this prediction,
suggesting that conclusive proof of the presence or absence of Sgr debris in the Solar neighborhood could prove a useful
tool for discriminating between models of the Galactic potential.
\end{description}

\acknowledgments
The authors would like to thank M.F. Skrutskie for helpful discussion, and 
D. Mart{\' i}nez-Delgado and M.A. G{\' o}mez-Flechoso for clarification on their satellite model
and for making available a pre-publication copy of their latest work.
SRM acknowledges support from Space Interferometry Mission Key Project
NASA/JPL contract 1228235, NSF grant AST-0307851,
a David and Lucile Packard Foundation Fellowship,
and the F.H. Levinson Fund of the Peninsula Community Foundation.
KVJ's contribution was supported through NASA grant NAG5-9064 and NSF CAREER award 
AST-0133617.

\clearpage
\begin{table}[h]
\tiny
\begin{tabular}{c|c|c|c||c|c}
\hline
Parameter/Property	& Description		& Value(s) tested		& Constrained$^{\it a}$ by:   & acceptable values$^{\it b}$ & value adopted\\
\hline
\hline
\multicolumn{5}{c}{\it Galactic parameters}\\
\hline
$d$			& scale length of		& 0 - 20 kpc		& 4,5	& 1 - 20 kpc & 13/12/11 kpc\\
			& the Galactic halo 		&		&	&&	\\
\hline
$q$			& flattening of the Galactic	& 0.8 - 1.4		& 6,7	& 0.8 - 1.4 & 0.90/1.0/1.25 \\
			& dark halo potential		&		&		&\\
\hline
$v_{\rm circ,\odot}$	  	& circular velocity      	& 180 - 240 km s$^{-1}$ & 4,5 	& 180 - 240 km s$^{-1}$ & 220 km s$^{-1}$\\
& at $R_\odot$ &&&\\
\hline
$\alpha$	  	                             & contribution of  disk	& 0.25 - 1.00 & 4,5 	& 0.25 - 1.00 & 1.00 \\
& to rotation curve &&&& \\
\hline
\multicolumn{5}{c}{\it Kinematical parameters}\\
\hline
			& tangential	& & & &\\
$v_{\rm tan}$ & velocity of Sgr & 200 - 400 km s$^{-1}$ & 3,4,5,7 & 230 - 330 km s$^{-1}$ & 280/270/254 km s$^{-1}$\\
\hline
$v_{\rm los, Sgr}$ 	& Sgr line of sight velocity   	& fixed	& 2	& 171 $\pm$ 1 km s$^{-1}$ & 171 km s$^{-1}$ \\
\hline
\multicolumn{5}{c}{\it Positional parameters}\\
\hline
$D_{\rm Sgr}$		& distance of Sgr		& 22 - 28 kpc		& 4,5	& 22 - 28 kpc & 24 kpc \\
               		& from the Sun 				&	&	&\\
\hline
$(l,b)_{\rm Sgr}$ 	& Galactic longitude and 	& fixed &  1	& --- & $(5.6^{\circ},-14.2^{\circ})$\\
			& latitude of the Sgr dwarf	&		&	&\\
\hline
$R_{\odot}$		& distance of the Sun from 	& 7.0 - 9.0 kpc & 4,5	& 7.0 - 9.0 kpc &7.0 kpc\\
			& the Galactic center				&		&	&\\
\hline

\multicolumn{5}{c}{\it Sagittarius dwarf parameters}\\
\hline
$M_{\rm Sgr}$           & present bound mass           & $6 \times 10^6$ - $3 \times 10^9 M_\odot$ & 8,9,10,11    
& 2 - 5$\times 10^8M_\odot$ &  $4 \times 10^8 M_\odot$\\
                        & of the Sgr dwarf$^{\it c}$             &               &                     &\\
\hline
\end{tabular}
\caption{Parameter space of Milky Way - Sgr models considered,
values quoted are for
each of three models of the Galactic potential ($q =$ 0.90/1.0/1.25 respectively).
Comments ---
{\it a.}: See \S2.2.
{\it b.}: Acceptable ranges of values are considerably smaller once fixed parameters are adopted.
{\it c.}: See Table \ref{mass.table} for further details.
}
\label{param.table}
\end{table}

\clearpage
\begin{table}[h]
\begin{tabular}{c|c|c|c}
\hline
Halo model & Constraint & Best-fit mass & Acceptable mass range\\
\hline
\hline
$q =$ 0.90 & $\sigma_{\rm v}$ & 2.3 $\times 10^8 M_{\odot}$ & 6.2 $\times 10^7 M_{\odot}$ - 5.0 $\times 10^8 M_{\odot}$\\
\hline
           & $\sigma_{Z_{Sgr,\odot}}$ & 3.8 $\times 10^8 M_{\odot}$ & 2.6 $\times 10^8 M_{\odot}$ - 5.3 $\times 10^8 M_{\odot}$\\
\hline
\hline
$q =$ 1.0 & $\sigma_{\rm v}$ & 2.8 $\times 10^8 M_{\odot}$ & 9.1 $\times 10^7 M_{\odot}$ - 5.3 $\times 10^8 M_{\odot}$\\
\hline
           & $\sigma_{Z_{Sgr,\odot}}$ & 3.7 $\times 10^8 M_{\odot}$ & 2.5 $\times 10^8 M_{\odot}$ - 5.4 $\times 10^8 M_{\odot}$\\
\hline
\hline
$q =$ 1.25 & $\sigma_{\rm v}$ & 4.8 $\times 10^8 M_{\odot}$ & 1.7 $\times 10^8 M_{\odot}$ - 8.6 $\times 10^8 M_{\odot}$\\
\hline
           & $\sigma_{Z_{Sgr,\odot}}$ & 3.8 $\times 10^8 M_{\odot}$ & 2.5 $\times 10^8 M_{\odot}$ - 5.5 $\times 10^8 M_{\odot}$\\
\hline
\end{tabular}
\caption{Acceptable values for the present-day bound mass of the Sgr dwarf ($M_{\rm Sgr}$) in each of our three halo models.}
\label{mass.table}
\end{table}

\clearpage


\clearpage

\clearpage

\clearpage


\clearpage

\clearpage

\clearpage

\clearpage

\clearpage

\clearpage

\clearpage

\clearpage

\clearpage


\clearpage
\begin{thebibliography}{99}

\bibitem[Binney \& Tremaine(1987)]{binney87} Binney, J.~\& 
Tremaine, S.\ 1987, Princeton, NJ, Princeton University Press, 1987, p. 429





\bibitem[Edelsohn \& Elmegreen (1997)]{EE97} Edelsohn, 
D. J. \& Elmegreen, B. G. 1997, \mnras, 290, 7

\bibitem[Eke, Navarro, \& Steinmetz(2001)]{eke01} Eke, V.~R., 
Navarro, J.~F., \& Steinmetz, M.\ 2001, \apj, 554, 114 

\bibitem[G{\' o}mez-Flechoso, Fux, \& Martinet(1999)]{GFM99} 
G{\' o}mez-Flechoso, M. A., Fux, R., \& Martinet, L. 1999, \aap, 347,
77

 
\bibitem[Helmi(2004)]{helmi04}
Helmi, A. 2004, \apjl, 610L, 97

\bibitem[Helmi \& White(1999)]{helmi99}
Helmi, A. \& White, S. D.~M. 1999,
\mnras, 307, 495

\bibitem[Helmi \& White (2001)]{HW01} Helmi, A. \& White, S.D.M. 2001, 
\mnras, 323, 529

\bibitem[Hernquist \& Ostriker(1992)]{hernquist92}  
Hernquist, L. \&  Ostriker,
J.P. 1992, \apj, 386, 375


\bibitem[Ibata, Gilmore, \& Irwin (1995)]{ibata95}
Ibata, R. A.,  Gilmore, G. \& Irwin, M. J. 1995, \mnras, 277, 781

\bibitem[Ibata, Gilmore, \& Irwin(1994)]{ibata94} Ibata, R.~A., 
Gilmore, G., \& Irwin, M.~J.\ 1994, \nat, 370, 194 

\bibitem[Ibata \& Lewis (1998)]{ibata98} Ibata, R.~A.~\& Lewis, 
G.~F.\ 1998, \apj, 500, 575 

\bibitem[Ibata et~al. (2001)]{ibata01}
Ibata, R.A., Lewis, G.~F., Irwin, M.J., Totten, E., \& Quinn, T 2001,
\apj, 551, 294

\bibitem[Ibata et al.(1997)]{ibata97} Ibata, R.~A., Wyse, 
R.~F.~G., Gilmore, G., Irwin, M.~J., \& Suntzeff, N.~B.\ 1997, \aj, 113, 634

\bibitem[Jiang \& Binney(2000)]{jiang00} Jiang, I.~\& Binney, 
J.\ 2000, \mnras, 314, 468 

\bibitem[Jing \& Suto(2002)]{jing02} Jing, Y.~P.~\& Suto, Y.\ 
2002, \apj, 574, 538 

\bibitem[Johnston (1998)]{johnston98}
Johnston, K.V. 1998, \apj, 495, 297

\bibitem[Johnston, Hernquist, \& Bolte (1996)]{johnston96} 
Johnston, K. V., Hernquist, L., \& Bolte, M. 1996, \apj, 465, 278

\bibitem[Johnston, Law, \& Majewski (2004)]{johnston04}
Johnston, K.~V., Law, D.~R. \& Majewski, S.~R. 2004, ApJ submitted, astro-ph/0407565 (``Paper III'')

\bibitem[Johnston et al. (1999)]{johnston99} Johnston, K.~V., 
Majewski, S.~R., Siegel, M.~H., Reid, I.~N., \& Kunkel, W.~E.\ 1999, \aj, 
118, 1719 

\bibitem[Johnston, Sackett, \& Bullock (2001)]{johnston01} Johnston, K.~V.,
Sackett, P.~B., \& Bullock, J.~S. 2001, \apj, 557, 137

\bibitem[Johnston, Spergel, \& Hernquist (1995)]{johnston95} 
Johnston, K.~V., Spergel, D.~N., \& Hernquist, L.\ 1995, \apj, 451, 598 

\bibitem[Kundu et al.\,(2002)]{kundu02} Kundu, A., Majewski, S.R., Rhee, J., Rocha-Pinto, H.J., 
Polak, A.A., Slesnick, C.L., Kunkel, W.E., Johnston, K.V., Patterson, R.J., Geisler, D., 
Gieren, W., Seguel, J., Smith, V.V., Palma, C., Arenas, J., Crane, J.D., \& Hummels, C.B. 2002, \apjl, 576, 125

\bibitem[Law et al.(2004)]{Law2004} Law, D. R.,
Majewski, S. R., Johnston, K. V., \& Skrutskie, M. F. 2004,
in Satellites and Tidal Streams, eds. F. Prada, D. Martinez-Delgado, T. Mahoney,
ASP Conf. Ser., {\it in press } (astro-ph/0309567)

\bibitem[Majewski et al. (2004b)]{Majewski2004b}
Majewski, S.R. et al. 2004b, in preparation (``Paper V'')

\bibitem[Majewski et al. (2004a)]{Majewski2004a}
Majewski, S. R., Kunkel, W.E., Law, D.R., Polak, A.A., Rocha-Pinto, H.J., Crane, J.D., Frinchaboy, P.M., Hummels, C.B.,
Johnston, K.V., Patterson, R.J., Rhee, J., Skrutskie, M.F. \& Weinberg, M.D. 2004a, \aj, 128, 245

\bibitem[Majewski et al.(1999)]{Maj1999} Majewski, S. R., 
Siegel, M. H., Kunkel, W. E., Reid, I. N., Johnston, K. V., Thompson, 
I. B., Landolt, A. U., \& Palma, C. 1999, \aj, 118, 1709 

\bibitem[Majewski et al. (2003)]{Majewski2003}
Majewski, S. R., Skrutskie, M.F., Weinberg, M.D. \& Ostheimer, J.C. 2003, \apj, 599, 1082 (``Paper I'')

\bibitem[Mart{\'{\i}}nez-Delgado et al.(2004)]{martinez04} Mart{\'{\i}}nez-Delgado, D.,
G{\'o}mez-Flechoso, M.~{\' A}., Aparicio, A., \& Carrera, R.\ 2004, \apj, 601, 242 

\bibitem[Mateo, Olszewski, \& Morrison(1998)]{mateo98} Mateo, 
M., Olszewski, E.~W., \& Morrison, H.~L.\ 1998, \apjl, 508, L55 


\bibitem[Miyamoto \& Nagai(1975)]{miyamoto75}
Miyamoto, M. \& Nagai, R. 1975, \pasj, 27, 533

\bibitem[Monaco et al. (2004)]{monaco04}
Monaco, L., Bellazzini, M., Ferraro, F.~R., \& Pancino, E. 2004, astro-ph/0406350

\bibitem[Navarro, Frenk, \& White(1996)]{NFW96} Navarro, 
J.~F., Frenk, C.~S., \& White, S.~D.~M.\ 1996, \apj, 462, 563 

\bibitem[Newberg et al.(2003)]{newberg03} Newberg, H.~J.~et al.\ 
2003, \apjl, 596, L191 




\bibitem[Plummer(1911)]{plummer11}
Plummer, H.C. 1911, \mnras, 71, 460


\bibitem[Totten \& Irwin (1998)]{totten98}
Totten, E.J. \& Irwin, M.J. 1998, \mnras, 294, 1

\bibitem[Tremaine(1993)]{tremaine93} Tremaine, S.\ 1993, AIP 
Conf.~Proc.~278: Back to the Galaxy, 599 

\bibitem[Velazquez \& White (1995)]{velazquez95}
Velazquez, H. \& White, S. D.~M. 1995,
\mnras, 275, 23


\bibitem[Zhao(1998)]{zhao98} Zhao, H.\ 1998, \apjl, 500, L149

\bibitem[Zhao (2004)]{zhao04}
Zhao, H. 2004, \mnras, 351, 891

\end{thebibliography}
\end{document}